\documentclass[journal]{IEEEtran}
\def\BibTeX{{\rm B\kern-.05em{\sc i\kern-.025em b}\kern-.08em
    T\kern-.1667em\lower.7ex\hbox{E}\kern-.125emX}}
\usepackage{amsmath,amssymb,amsfonts}
\usepackage{algorithmic}
\usepackage[ruled]{algorithm2e}
\usepackage{subfigure}
\usepackage{array}
\usepackage{textcomp}
\usepackage{stfloats}
\usepackage{url}
\usepackage{verbatim}
\usepackage{graphicx}
\usepackage{cite}
\usepackage{booktabs}
\usepackage{makecell}
\usepackage{multirow}
\usepackage{balance}
\usepackage{amssymb}
\usepackage{bbding}
\usepackage{pifont}
\usepackage[table]{xcolor}
\usepackage{tikz}
\usetikzlibrary {spy}
\usepackage{times}
\usepackage{mathptmx}

\usepackage{newtxtext}
\makeatletter  
\newcommand{\Rmnum}[1]{\expandafter\@slowromancap\romannumeral #1@}
\makeatother

\usepackage{hyperref}
\hypersetup{ 
    colorlinks=true, 
    linkcolor=black, 
    anchorcolor=black, 
    urlcolor=black,
    citecolor=black} 

\usepackage{orcidlink} 
\begin{document}

\renewcommand{\algorithmicrequire}{\textbf{Input:}} 
\renewcommand{\algorithmicensure}{\textbf{Output:}}
\title{Dreamer: Dual-RIS-aided Imager in Complementary Modes\\

\thanks{This work is supported by the National Natural Science Foundation of China under Grant No.12141107, and the Interdisciplinary Research Program of HUST (2023JCYJ012). (Corresponding author: Zenan Ling)}
\thanks{Fuhai Wang is with the School of Electronic Information and Communications, Huazhong University of Science and Technology, Wuhan 430074, China, and also with the Institute of Artificial Intelligence, Huazhong University of Science and Technology, Wuhan 430074, China.}
\thanks{Yunlong Huang, Rujing Xiong, Zhe Li, Chun Wang, Tiebin Mi, Robert Caiming Qiu, and Zenan Ling are with the School of Electronic Information and Communications, Huazhong University of Science and Technology, Wuhan 430074, China (email:lingzenan@hust.edu.cn; caiming@hust.edu.cn).}
\thanks{Zhanbo Feng is with the Department of Computer Science and Engineering, Shanghai Jiao Tong University, Shanghai 200030, China.}
}

\author{\IEEEauthorblockN{Fuhai Wang, 
Yunlong Huang, 
Zhanbo Feng, 
Rujing Xiong, 
Zhe Li, 
Chun Wang,~\IEEEmembership{Student Member,~IEEE,} 
Tiebin Mi,~\IEEEmembership{Member,~IEEE,} 
Robert Caiming Qiu,~\IEEEmembership{Fellow,~IEEE,} 
and Zenan Ling,~\IEEEmembership{Member,~IEEE,}} \\
}

\maketitle
\begin{abstract} 
Reconfigurable intelligent surfaces (RISs) have emerged as a promising auxiliary technology for radio frequency imaging. However, existing works face challenges of faint and intricate back-scattered waves and the restricted field-of-view (FoV), both resulting from complex target structures and a limited number of antennas. The synergistic benefits of multi-RIS-aided imaging hold promise for addressing these challenges. Here, we propose a dual-RIS-aided imaging system, Dreamer, which operates collaboratively in complementary modes (reflection-mode and transmission-mode). Dreamer significantly expands the FoV and enhances perception by deploying dual-RIS across various spatial and measurement patterns. Specifically, we perform a fine-grained analysis of how radio-frequency (RF) signals encode scene information in the scattered object modeling. Based on this modeling, we design illumination strategies to balance spatial resolution and observation scale, and implement a prototype system in a typical indoor environment. Moreover, we design a novel artificial neural network with a CNN-external-attention mechanism to \textbf{translate} RF signals into high-resolution images of human contours. Our approach achieves an impressive SSIM score exceeding $0.83$, validating its effectiveness in broadening perception modes and enhancing imaging capabilities. The code to reproduce our results is available at {\url{https://github.com/fuhaiwang/Dreamer}}.
\end{abstract}


\section{Introduction} 
\IEEEPARstart{I}n contemporary society, the prevalence of radio-frequency (RF) imagers in the electromagnetic (EM) sensing domain is on the rise. RF imaging technology demonstrates its capability to generate high-resolution images under varying lighting conditions and weather fluctuations, supporting applications such as environmental monitoring~\cite{HIMEUR202244, depatla2017robotic}, medical monitoring~\cite{rosen2002applications, chandra2015opportunities}, autonomous driving~\cite{orr2021coherent} and security inspection~\cite{li2024high, adib2013see}. Conventional RF imaging systems depend on either frequency diversity~\cite{9257482} or spatial diversity~\cite{li2024high,li2008signal} to achieve precise sensing. Notably, specific signal modulation techniques broaden the bandwidth of imaging signals to improve the degree of freedom through frequency diversity, such as ultra-wideband (UWB) technology~\cite{ahmed2021uwb}, linear frequency modulation (LFM)~\cite{10230998}, and orthogonal frequency-division multiplexing (OFDM)~\cite{9076313}. Moreover, implementing antenna diversity techniques, such as multiple input multiple output (MIMO) antennas~\cite{Adib2015Capturing}, can achieve super-resolution imaging in the angular domain and provide significant spatial diversity. Although expanding the signal bandwidth and increasing the number of RF antenna units can improve image resolution, it also introduces challenges such as higher power consumption, increased complexity, and higher costs in transceiver systems.

Reconfigurable intelligent surfaces~\cite{cui2014coding,8796365}, as an active member of the metamaterial family, offer characteristics such as low power consumption, cost-effectiveness, and easy deployment. The integration of RIS-based passive antennas into computational imaging (CI) has provided a fresh approach to broaden imaging's degrees of freedom~\cite{wang2016single, gollub2017large, Lipworth:15, li2019machine}. Specifically, RIS-based solutions offer pattern diversity~\cite{hunt2013metamaterial} in measurements by adjusting the phase shifts of these elements to reconfigure EM waves and fields, complementing the current schemes based on frequency diversity and spatial diversity. Recent studies on RIS-aided RF imaging systems highlight the potential of RIS technology to dynamically tailor the radio environment~\cite{cui2020wireless,huang2021forgery,he2022high,li2019machine,li2019intelligent,zhao2023intelligent}, which enhances imaging capabilities. In real-world deployment, a RIS generally serves the transmitter (Tx) to scatter the incoming EM signal and illuminate the object. The reflected signal, encoding the target information, is then received by the receiver (Rx). Here, RISs provide a uniquely flexible platform for exploiting CI capabilities, multiplexing the reflectivity map of human targets. 

The RIS-aided reflection-mode imaging system offers an approach, particularly for targets with high reflection coefficients, to enhance the multiplexing capacity of spatial information~\cite{sleasman2020implementation,saigre2022intelligent,imani2020review,yurduseven2015resolution,boyarsky2018single,imani2018two}. However, intricate back-scattered waves~\cite{gennarelli2014effect, gollub2017large} and a limited field-of-view~\cite{li2019machine} often result in blurred and incomplete images of objects. Firstly, the complexity of target structures and surface roughness relative to the illumination wavelength often results in multiple scattering, generating faint and intricate back-scattered waves~\cite{maitre2013processing, bensky2019short, gollub2017large}. Extracting information for reconstruction algorithms~\cite{gonzalez2013use,amineh2014microwave} from such complex, low signal-to-noise ratio reflected waves is challenging, especially from high-loss objects with low reflection coefficients~\cite{gennarelli2014effect}. Additionally, the RF imaging system requires numerous transceiver antennas to achieve a synthetic aperture, thereby enhancing the imaging space (a wider FoV) and enabling high-resolution spatial measurements~\cite{wilson2010radio,li2019machine}. However, these requirements lead to complex system designs and extremely expensive hardware. 

\begin{figure}
    \centerline{\includegraphics[width=0.9\columnwidth]{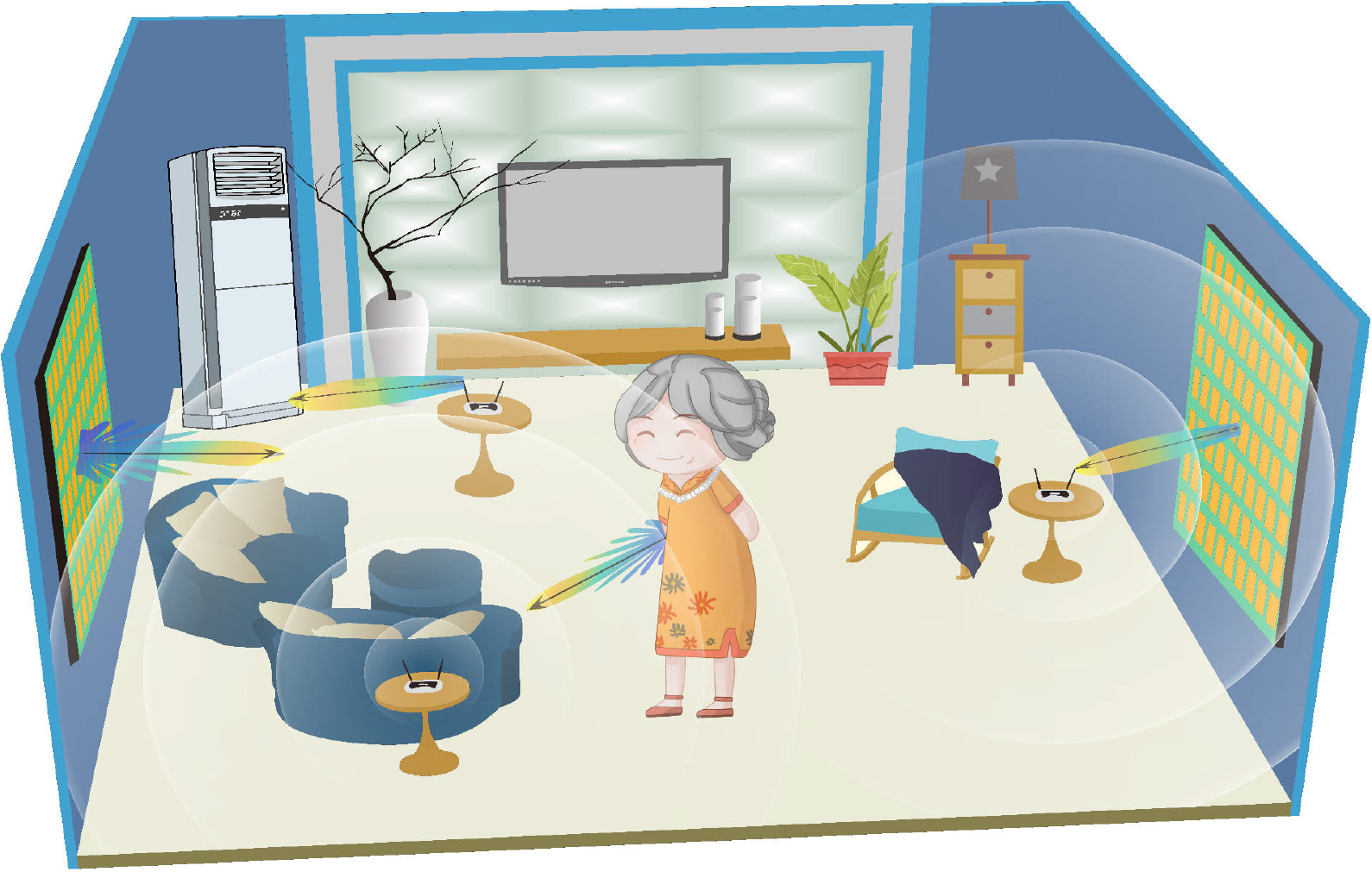}} 
    \caption{A scenario for smart and cost-effective wireless imaging of the elderly indoors involves installing dual-RIS integrated into the walls of living rooms or hallways. This system serves as a complementary method to cameras in scenarios with low light, line-of-sight obstructions, and privacy concerns.} 
    \label{Scene diagram} 
\end{figure}

To this end, we introduce a novel \textbf{d}ual-\textbf{re}configurable intelligent surface-\textbf{a}ided i\textbf{m}ag\textbf{er} (\textbf{Dreamer}) system. We clarify a hybrid transmission–reflection-mode~\cite{mervcep2019transmission} in RIS-aided imaging systems. The hybrid-mode (complementary mode), offering a full-view perspective, captures the unique reflectance and transmittance properties of the target, which we visualize through a contrast map depicting differences the target and the background. Additionally, the RIS functions as a large antenna aperture of the passive relay ``antenna'' array. When RISs with large apertures dynamically serve both transmitters and receivers, expanding the imaging system's precise range and FoV becomes feasible. This means that Dreamer exhibits a high degree of freedom and a modified FoV similar to a bistatic radar configuration~\cite{buzzi2022foundations}. Fig.~\ref{Scene diagram} shows the scenario of Dreamer, which monitors individuals indoors in a smart and cost-effective manner by integrating dual-RIS into the walls. Compared to large-scale transceiver antennas~\cite{wang2015industrial}, the feasible and economical dual-RIS-aided approach indicates a promising future for wireless imaging. 

Dreamer is an end-to-end RF imaging system assisted by a forward RIS and a backward RIS. Dreamer collaboratively controls the measurement channels in a complementary mode. The novelty of Dreamer lies in the integration of both the signal transmitted through the space of interest (SoI) and the signal reflected from the target for imaging. To improve pattern diversity, we design illumination strategies which enable fine-grained RF sensing with a high degree of freedom. To reconstruct the human contour from complicated RF signals, we further develop a innovative CNN-external-attention-based model, which plays a crucial role in extracting the latent representation of target-related signals in a complex electromagnetic environment. As a result, Dreamer demonstrates the synergistic and complementary value of the multi-modal combination, achieving a SSIM score exceeding $0.83$.

\subsection{Related Work} \label{Related work}
\subsubsection{Metasurface-Based Imaging} 
Metasurfaces, as proposed and experimentally demonstrated in previous works~\cite{sleasman2015dynamic}, offer a promising approach for dynamically altering radiation patterns using switchable components integrated into metamaterial elements within a waveguide. Metasurfaces enable the manipulation of electromagnetic responses and the encoding of scene information through computational imaging schemes~\cite{imani2020review,yurduseven2015resolution,boyarsky2018single,imani2018two}. RISs, as part of the metamaterial family, share similarities with metasurfaces in dynamically adjusting parameters to control electromagnetic responses. RISs function as passive relay antenna arrays in integrated sensing and communication systems~\cite{huang2023joint}. By adjusting phase, amplitude, and frequency, RISs manipulate the wireless environment and acquire scene information through spatially diverse patterns. 

Future research directions may involve exploring how different combinations of antennas on RISs can enhance sensing performance in RIS-based imaging systems~\cite{lan2020wireless}. multi-RIS-based imaging systems could involve investigating optimal antenna configurations and tuning strategies to improve imaging resolution, coverage, and efficiency. Additionally, exploring synergies between metasurfaces and RISs could lead to innovative approaches for controlling electromagnetic waves and enhancing imaging capabilities in various applications, including wireless communication, sensing, and imaging.

\subsubsection{Computational Microwave Imaging} 
In sensing-based imaging, the primary goal is to estimate the material properties of the environment by analyzing the signals received from various sensing modalities. The process involves encoding scene information through different illumination patterns or measurement modes, which play a pivotal role in computational electromagnetic imaging~\cite{boyarsky2017synthetic,sleasman2017experimental}. 

Conventional computational microwave imaging systems have advanced significantly by integrating non-intuitive hardware designs with CI strategies~\cite{saigre2022intelligent}. The relationship between the received data $y$, the scene $x$ (containing $N$ pixels), and the imaging system or measurement process $\mathcal{C}$ can be mathematically represented as $y=\mathcal{C}(x)+n$, where $n$ represents noise. The imaging task is to estimate the image $x$ from the received data $y$ and the known measurement process $\mathcal{C}$, essentially solving an inverse scattering problem (ISP)~\cite{karl2023foundations,chen2018computational}. Currently, \textsl{non-iterative methods} and \textsl{AI-based techniques} are considered two main approaches in the field of ISPs.

Non-iterative inversion methods aim to linearize ISPs under specific conditions~\cite{6490334,5592613,deshmukh2022physics,saliani2022space,8363016}, such as the Born approximation (BA) or the Rytov approximation method~\cite{chen2018computational}. These approximations simplify the complex relationship between the received signals and the scattering objects, making the inversion process more tractable. However, these methods often assume free-space conditions and do not account for the multipath effects present in complex indoor environments. This limitation poses challenges when modeling the relationship between received signals and scattering objects in an indoor setting with multiple RISs.

AI-based reconstruction techniques have garnered significant attention for their potential to improve image quality in complex environments~\cite{wang2022intelligent,hu2022metasketch,9392006,lan2020wireless,9825738}. These techniques leverage the powerful feature extraction and learning capabilities of AI models~\cite{9785455,wei2018deep}, which have been shown to surpass traditional image reconstruction methods. Over recent years, AI-based models have been applied to various downstream tasks in sensing systems, such as action recognition~\cite{wangmulti}, gesture recognition~\cite{li2019intelligent}, breath monitoring~\cite{hu2020reconfigurable}, and other intelligent sensing applications~\cite{zhao2023intelligent}. Despite these advancements, designing an AI-based model that effectively reconstructs images from RF signals remains an open challenge. The complexity arises from the need to accurately model the multipath effects and interactions in the indoor environment, which are crucial for high-resolution imaging. 

\subsubsection{Illumination Strategies of Metasurface} \label{Illumination Strategies of Metasurface}
Illumination patterns are pivotal in determining the effectiveness and efficiency of imaging systems. Two primary illumination modes commonly used are the \textsl{spotlight mode} (diffraction-limited resolution)~\cite{yurduseven2017millimeter} and the \textsl{multiplex measurement mode}~\cite{van2019coding}. 

The spotlight mode establishes a one-to-one mapping from scene voxels to reconstructed image pixels. This mode utilizes the metasurface to form a pencil beam that focuses on a specific pixel within the detected target, as demonstrated in previous studies. The focused beamforming mode enhances the signal-to-noise ratio (SNR) without sacrificing resolution, which is advantageous for achieving high-quality imaging of small areas. However, this mode becomes inefficient when imaging larger SoI due to the substantial increase in the number of measurements required, leading to significant measurement inefficiency~\cite{duarte2008single}.

Multiplex measurement imaging methods have been proposed for electromagnetic imaging to address the inefficiency of the spotlight mode. This approach leverages spatial voxel information reuse (multiplex advantage)~\cite{fellgett1949ultimate} and the various illumination patterns provided by the RIS to achieve multiplex imaging of scene pixels. The simplest multiplex method involves random mode illumination, but more sophisticated, carefully designed schemes have also been developed~\cite{li2019intelligent,boyarsky2017synthetic}. Multiplex modes, which can be seen as complex linear combinations of different illumination patterns, reduce the number of required samples and accelerate the imaging process. This approach leverages the theoretical advantages of compressed sensing, including universality, robustness, and sparsity~\cite{duarte2008single}. However, the multiplexing effect may introduce noise and degrade the quality of the imaging, necessitating a balance between speed and image fidelity. 

\subsection{Contributions and Paper Organization}

The main contributions are summarized as follows. 
\begin{enumerate}  
\item  We introduce a transmission–reflection-mode RISs-aided RF imaging system, assisted by a forward RIS and a backward RIS. Additionally, we propose illumination strategies to further increase pattern diversity. By leveraging the radiation pattern diversities of the two RISs, we improve imaging accuracy and coverage area. 

\item  We theoretically model the transmitted and reflected signals aided by the dual-RIS setup using inverse scattering theory. Building upon this theoretical insight, we design a novel deep learning network, named \textbf{PSD2ImageNet}, to solve inverse scattering problems by reconstructing indoor images from received signals. Specifically, a hybrid CNN-external-attention architecture is proposed to extract the latent representation of the object, surpassing conventional convolutional methods. 

\item  To the best of our knowledge, we are the first to implement a proof-of-concept dual-RIS-aided imaging system. Extensive experiments demonstrate that our system provides high-resolution human imaging, with the average SSIM between the imaging results and the real human contour exceeds 0.83, indicating strong fidelity. Furthermore, the experimental results highlight the superiority of both the dual-RIS performance and the employed reconstruction algorithm. 
\end{enumerate}  

The paper is organized as follows. Section \ref{Related work} discusses related work in the field. Section \ref{System Model section} presents the signal modeling of the dual RIS Imager. Section \ref{System Design} outlines the system design, including the system workflow, illumination strategies, and reconstruction algorithm. Section \ref{Experiment} elaborates on the experiments and the implementation of the imaging system. Evaluation outcomes are discussed in Section \ref{S:Performance analysis}. Finally, conclusions are drawn in Section \ref{S:Conclusion}. 

\begin{figure}
    \centerline{\includegraphics[width=1\columnwidth]{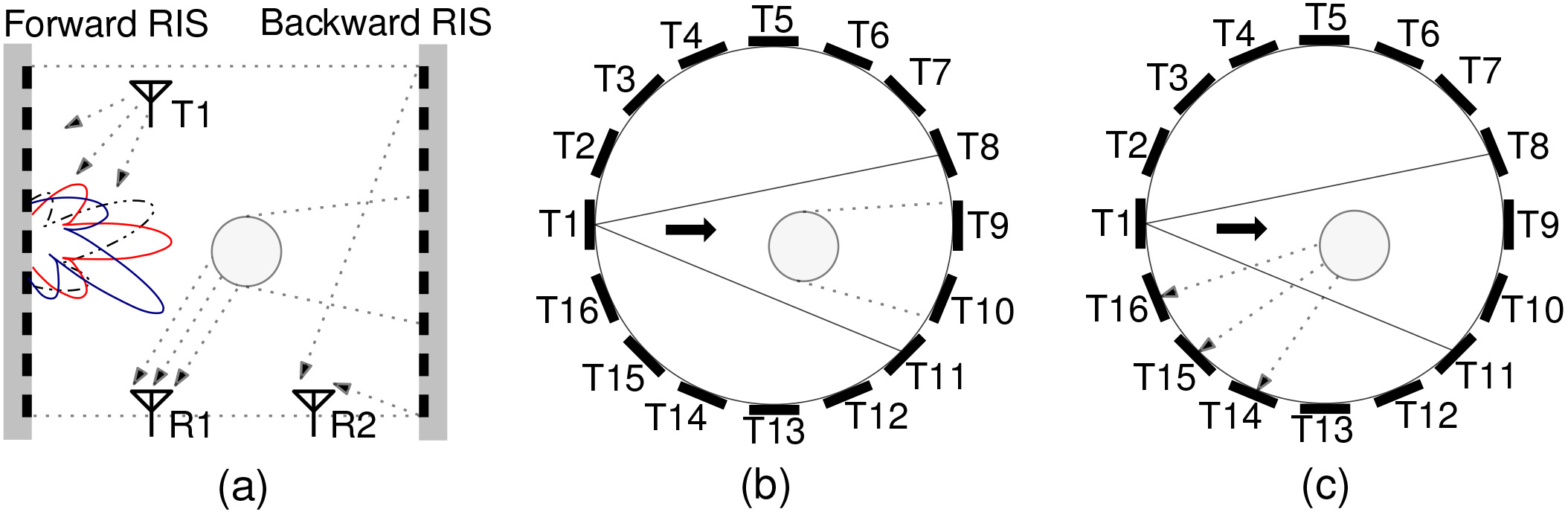}}
    \caption{({{a}}) The proposed dual-RIS-aided imaging system establishes two operational modes for the receivers: transmission-mode and reflection-mode. These modes resemble those in transmission and reflection tomography systems~\cite{mervcep2019transmission}. Traditional systems rely on omnidirectional collection of radio frequency signals through transmitting and receiving sensors. In contrast, the proposed imaging system employs the reconfigurable of multi-degree-of-freedom data collection through dual-RIS, thus avoiding the need for numerous and costly transceiver devices. Both reflection-mode and transmission-mode are complementarily implemented in the proposed imaging system. Operation of traditional ({{b}}) transmission and {{(c)}} reflection tomography systems with a large number of transceivers. In tomography systems, reflection devices operate by sending an electromagnetic pulse, which is reflected at an interface with different electrical impendence before being detected by the transducers. Transmission systems propagate an electromagnetic wave through a material and gather information on the systems by the time and amplitude at which the pulse is received by transducers at different radial locations.} 
    \label{tomography_RIS}
\end{figure}

\section{System Model} \label{System Model section}
\subsection{Deployment Principle of the Dual-RIS} 
Our system consists of a Tx, two Rxs, and two RISs. The schematic diagram of the proposed dual-RIS-aided imaging system is depicted in Fig.~\ref{tomography_RIS}{a}. By deploying two RISs, we establish two operational modes, transmission-mode and reflection-mode, for the receivers. These modes are analogous to two prevalent tomography imaging techniques in RF imaging~\cite{wang2015industrial}. The basic schematic principle highlights the operational differences between the transmission tomography system and the reflection one, as shown in Fig.~\ref{tomography_RIS}{b} and Fig.~\ref{tomography_RIS}{c}. While the target in RF systems behaves as an impenetrable scatterer, belonging to the obstacle problem~\cite{chen2018computational}, from the perspective of the SoI, it resembles the principles of traditional ultrasound tomography~\cite{tan2019ultrasonic, malik2016objective, yuan2023noninvasive}. For the obstacle problem, we utilize dual-RIS to improve the traditional approach to RF imaging and reconstruct images using transmission–reflection-mode signals in a complementary manner. 

\subsection{Signal Modeling} \label{Received Signal Modeling} 
Human imaging by the system can be framed as an obstacle problem~\cite{chen2018computational,colton1998inverse}, as well as a lossy dielectric object problem~\cite{gennarelli2014effect, sasaki2022measurement}. While it is feasible for the system to delineate the target contour by transmitting the signal, back-scattering signals may be very weak due to the low signal-to-noise ratio. Hence, the possibility to reconstruct objects based on both transmitted field and reflected fields becomes attractive. To describe the received signals, we utilize the Born approximation~\cite{chen2018computational, lipworth2013metamaterial}, providing an intuitive and concise depiction of how the behaviors of RIS radiation influence the received signals and object information.

In a homogeneous dielectric background with dielectric constant $\epsilon_0 $ and magnetic permeability $\mu_0 $, a nonmagnetic scatterer (the object) with relative permittivity $\epsilon_\mathrm{r}(\mathbf{r}) $ is located in the SoI $ D \subset R^3$ and is illuminated from different angles by time-harmonic electromagnetic waves. The SoI comprises a total number of $M$ point-like scatterers, which is discretized into $Q$ cells ($ M \ll Q$). The centers of these cells are at $\mathbf{r}_n$ ($n = 1, 2, \dots, Q$). The area and relative permittivity of the $n$-th cell are denoted by $A_n$ and $\epsilon_\mathrm{r}(\mathbf{r}_n)$, respectively. Each point-like scatterer is assumed to occupy a cell with good approximation.

\begin{figure*}
    \begin{equation} \label{signal_y1}
        \begin{aligned}
            y_1(\mathbf{r}_\mathrm{R_1};\omega,\mathcal{A}) \approx & \underbrace{{s(\omega)} {\sum_n^{Q} k_0^2(\omega) A_n g(\mathbf{r}_\mathrm{R_1},\mathbf{r}_n;\omega) \chi(\mathbf{r}_n) }
            { \times \sum_{i}^{K} k_0^2(\omega) \Upsilon^{\mathcal{A}}_i(\omega) g(\mathbf{r}_i,\mathbf{r}_n;\omega) g(\mathbf{r}_\mathrm{T},\mathbf{r}_i;\omega) }}_{ \text{Tx}\rightarrow \text{Forward RIS} \rightarrow \text{Target} \rightarrow \text{Rx}_1} + \underbrace{s(\omega)g(\mathbf{r}_\mathrm{T},\mathbf{r}_\mathrm{R_1};\omega)}_{ \text{Tx}\rightarrow \text{Rx}_1} + \kappa_1(\omega)
        \end{aligned} 
    \end{equation} 
\end{figure*} 

\begin{figure*}
    \begin{equation} \label{signal_y2}
        \begin{aligned}
        y_2(\mathbf{r}_\mathrm{R_2};\omega,\mathcal{A},\mathcal{B}) \approx &  \underbrace{{s(\omega)} {\sum_{j}^{K} k_0^2(\omega) \Upsilon^{\mathcal{B}}_j(\omega) g(\mathbf{r}_\mathrm{R_2},\mathbf{r}_j;\omega)} { \sum_{i}^{K} k_0^2(\omega) \Upsilon^{\mathcal{A}}_i(\omega) g(\mathbf{r}_i,\mathbf{r}_j;\omega) } { {\mu (\mathbf{r}_i, \mathbf{r}_j, \chi)} g(\mathbf{r}_\mathrm{T},\mathbf{r}_i;\omega) }}_{ \text{Tx}\rightarrow \text{Forward RIS} \rightarrow \text{Backward RIS} \rightarrow \text{Rx}_2}  \\
        & + \underbrace{ {s(\omega)} {\sum_{j}^{K} k_0^2(\omega) \Upsilon^{\mathcal{B}}_j(\omega) g(\mathbf{r}_\mathrm{R_2},\mathbf{r}_j;\omega) } {\sum_n^{Q} k_0^2(\omega) A_n g(\mathbf{r}_j,\mathbf{r}_n;\omega) \chi(\mathbf{r}_n) } { \sum_{i}^{K} k_0^2(\omega) \Upsilon^{\mathcal{A}}_i(\omega) g(\mathbf{r}_i,\mathbf{r}_n;\omega) g(\mathbf{r}_\mathrm{T},\mathbf{r}_i;\omega) }}_{ \text{Tx}\rightarrow \text{Forward RIS} \rightarrow \text{Target} \rightarrow \text{Backward RIS}\rightarrow \text{Rx}_2} \\
        & + \kappa_2(\omega),
        \end{aligned}
    \end{equation}
\end{figure*}

The total number of units for each RIS is denoted by $K$. The Tx horn antenna, fixed at $\mathbf{r}_\mathrm{T}$, emits an RF signal represented by $s(\omega)$, where $\omega$ is the angular frequency. $\text{Rx}_1$ is located at $\mathbf{r}_\mathrm{R_1}$, and $\text{Rx}_2$ is located at $\mathbf{r}_\mathrm{R_2}$. Thus, the forward models of the echo signal $y_1$ received by $\text{Rx}_1$ are to estimate the object function from the measured scattered fields. $y_1(\mathbf{r}_\mathrm{R_1};\omega,\mathcal{A})$ is defined by Eq.~\eqref{signal_y1}, with the symbols listed below.
\begin{itemize}
    \item $\mathcal{A}$ is the coding illumination patterns of forward RIS.
    \item $\Upsilon^{\mathcal{A}}_i$ is the reflection coefficient of the $i$-th unit of the forward RIS. The reflection coefficient is mainly related to the unit area, relative permittivity, and the control coding illumination patterns.
    \item $k_0(\omega) = \omega \sqrt{\mu_0 \epsilon_0}$ is the wavenumber of the homogeneous medium background in the angular frequency $\omega$. 
    \item $g\left(\mathbf{r}, \mathbf{r}^{\prime} ; \omega\right) = \frac{{{e^{i{k_0(\omega)}\left| {{\bf{r}} - {\bf{r}}'} \right|}}}}{{4\pi \left| {{\bf{r}} - {\bf{r}}'} \right|}}$ is the scaler Green's function in the angular frequency $\omega$, which is treated as an impulse response function.
    \item $\chi (\mathbf{r}) = \epsilon_\mathrm{r}(\mathbf{r})-1$ is the contrast which is determined by the object in the SoI. If there is no point-like scatterer in the cell, $\chi (\mathbf{r}) = 0$.
    \item $\kappa_1(\omega)$ accounts for disturbances from inband inferences, environment clutters, system noise, and others. 
 \end{itemize} 
In Eq.~\eqref{signal_y1}, the first term represents the reflection signal that is modulated by the forward RIS and further reflected by the human body before being collected. The second term represents the signal transmitted from the Tx to the $\text{Rx}_1$. 

The transmitted signal $y_2$ received by $\text{Rx}_2$ is approximated as Eq.~\eqref{signal_y2}, where the symbols are listed below. 
\begin{itemize}
    \item $\mathcal{B}$ is the coding illumination patterns of backward RIS.
    \item $\Upsilon^{\mathcal{B}}_j$ is the reflection coefficient of the $j$-th cell of the backward RIS. The reflection coefficient is mainly related to the unit area, relative permittivity, and the control coding illumination patterns.
    \item $\mu(\mathbf{r}_i,\mathbf{r}_j, \chi)$ serves as a transmission coefficient flag. If there is a scatterer between $\mathbf{r}_i$ and $\mathbf{r}_j$, the transmission coefficient flag $\mu(\mathbf{r}_i,\mathbf{r}_j, \chi) = 0$~\cite{pathak2021electromagnetic}. Otherwise, $\mu(\mathbf{r}_i,\mathbf{r}_j, \chi) = 1$. The flag indicates whether there is a scatterer between the $i$-th unit of the forward RIS and the $j$-th unit of the backward RIS. 
    \item $\kappa_2(\omega)$ accounts for disturbances from inband inferences, environment clutters, system noise, and others. 
\end{itemize} 

In Eq.~\eqref{signal_y2}, the first term indicates the transmitted signal modulated by the forward RIS and directly aggregated to the receiver through the backward RIS. The transmitted signal term can effectively indicate the presence or absence of the target through $\mu(\mathbf{r}_i,\mathbf{r}_j, \chi)$, similar to transmission tomography systems~\cite{wang2015industrial}. In the obstacle problem~\cite{chen2018computational,colton1998inverse}, if there is a scatterer in the propagation path, the signal will be blocked. The backward RIS receives only the scattered waves from the target, i.e., the second term of Eq.~\eqref{signal_y2}. This term indicates that the signal is modulated by the forward RIS, scattered by the target in the SoI, and modulated by backward RIS before being received by $\text{Rx}_2$. Since the backward RIS is placed behind the target, most of the scattered signals cannot illuminate it. The transmitted signal dominates in the signal $y_2$. The two acquired signals form two complementary signal modes. 

\subsection{Problem Description}
Each received signal equation contains information $\chi (\mathbf{r})$ about the object. In other words, the contrast vector $\bf{\chi(\mathbf{r})}$ carries the information of human postures. On the one hand, from the system design level, our objective is to amplify the terms in the received signal that are related to the target through the design of the illumination strategy. On the other hand,  from a practical task perspective, we aim to retrieve the reflection coefficients of the object from the received signal. To obtain the object information $\chi$ through the scattered signal and the transmitted signal described in the above equations, we must solve such a non-linear inverse problem. 

Given the open nature of this problem, achieving precise inverse scattering solutions through conventional theoretical modeling is challenging. Deep learning models can leverage powerful feature extraction and learning capabilities to address the non-linear ISP. Our goal is to design a novel inverse scattering imaging neural network capable of retrieving the reflection coefficients $\chi(\mathbf{r})$ from the received data and subsequently imaging the object. 

\section{System Design} \label{System Design} 
\begin{figure}
    \centerline{\includegraphics[width=1\columnwidth]{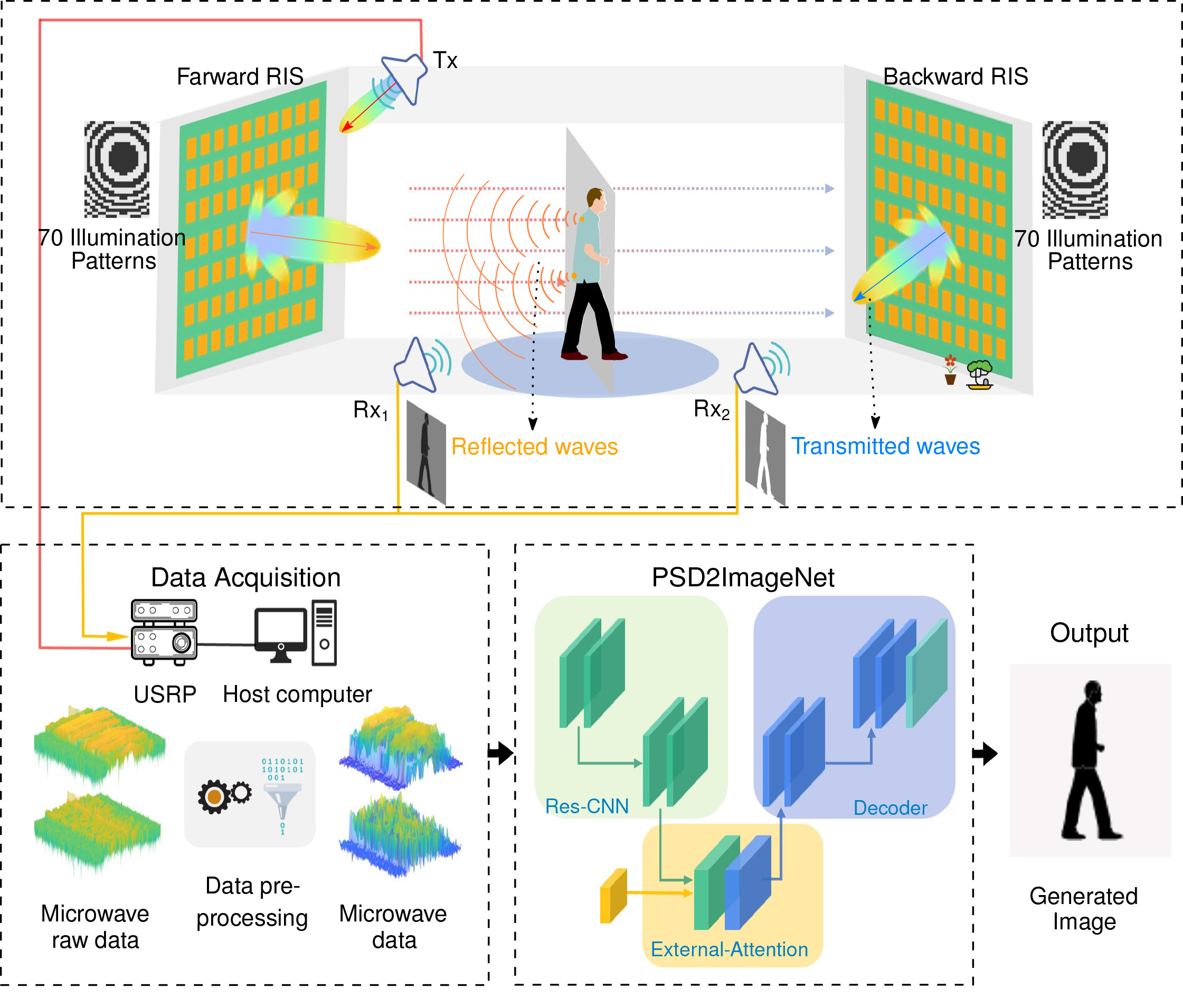}}
    \caption{Schematic of the dual-RIS-aided imager framework. The framework of the dual-RIS-aided imager comprises three parts: hardware deployment, data acquisition strategy and image reconstruction. In the hardware deployment part, the forward RIS collaborates with the Tx to shape radio propagation, while $\text{Rx}_1$ with a horn antenna hears the echoes reflected from objects. The backward RIS complements the forward RIS by capturing unblocked and unreflected signals (transmitted, diffracted, and scattered signals) as the electromagnetic signal traverses the SoI. The imaging system simultaneously operates in two complementary imaging modes. In the data acquisition strategy part, the system gathers microwave data using two RISs operating in different illumination patterns to serve both Tx and Rx. In the image reconstruction part, Dreamer feeds the preprocessed microwave data into the pre-trained image reconstruction network for image generation.} 
    \label{scene schematic}
\end{figure}

The framework used in this paper consists of three parts: hardware deployment, data acquisition strategy and image reconstruction, as shown in Fig.~\ref{scene schematic}. 

In the hardware deployment part, the fundamental hardware design involves deploying two RISs to serve the Tx and the Rx operating at $5.8$ GHz, respectively. One of the RISs, named the forward RIS, collaborates with the Tx to shape radio propagation and imaging objects in the SoI. The SoI is located in front of the forward RIS, and the receiver ($\text{Rx}_1$) with a horn antenna captures the echoes reflected from objects. The echo path is represented as ``Tx$\rightarrow$Forward RIS$\rightarrow$Target$\rightarrow$$\text{Rx}_1$.'' The other RIS referred to as the backward RIS, complements the forward RIS by capturing remaining signals that are not obstructed or reflected by objects as the electromagnetic signal traverses the SoI. The receiver ($\text{Rx}_2$) with a horn antenna captures the transmitted signals beamformed by the backward RIS. The transmission-mode path is represented as ``Tx$\rightarrow$Forward RIS$\rightarrow$SoI$\rightarrow$Backward RIS$\rightarrow$$\text{Rx}_2$.'' The manner of signal acquisition by $\text{Rx}_1$ and $\text{Rx}_2$ is ensured to resemble the reflection-mode and transmission-mode in tomography systems~\cite{gennarelli2014effect,anastasio2000new}, respectively. 

In the data acquisition strategy part, we further utilize the reconfigurable capability of RIS to design three types of illumination strategies, resulting in diverse radiation patterns. Two commonly used illumination patterns are the diffraction-limited spotlight mode and the multiplex measurement mode. The former establishes a one-to-one mapping from scene voxels to reconstructed image pixels, with the metasurface~\cite{sleasman2015dynamic} forming a pencil beam that focuses on a specific pixel in the detected target. By leveraging spatial voxel information and employing diverse illumination patterns, we achieve efficient multiplex imaging of scene pixels, known as the multiplex advantage~\cite{fellgett1949ultimate}. These various multiplex measurement modes generate linear combinations, effectively reducing the number of samples and accelerating the imaging process. Moreover, jointly controlling the illumination strategies between the forward RIS and the backward RIS brings synergistic benefits. 

In the image reconstruction part, we design a novel CNN-external-attention-based model to reconstruct the human contour images from the RF data. To provide an intuitive and concise understanding of how the radiation behaviors of RIS influence the received signals and object information, we model both the reflected and transmitted signals. Details information on signal modeling and the problem description can be found in the Methods section. Our neural network design incorporates an essential external attention module, crucial for addressing the ISP within the complexities of the electromagnetic environment. The model is trained in a supervised manner, with the ground truth being the contour data segmented from the image data which is collected synchronously by the camera during RF data acquisition. 

\subsection{Design of Illumination Strategies} \label{Illumination Strategies}
The reflection-mode and transmission-mode signals are received by $\text{Rx}_1$ and $\text{Rx}_2$, respectively. The first term (the ``Tx$\rightarrow$Forward RIS$\rightarrow$Target$\rightarrow$$\text{Rx}_1$'' link) in Eq.~\eqref{signal_y1} and the first two-term (the ``Tx$\rightarrow$Forward RIS$\rightarrow$Backward RIS$\rightarrow$$\text{Rx}_2$'' link and the ``Tx$\rightarrow$Forward RIS$\rightarrow$Target$\rightarrow$Backward RIS $\rightarrow$$\text{Rx}_2$'' link) in Eq.~\eqref{signal_y2} demonstrate the necessity of enhancing their weight to improve the resolution of the object's imaging and the signal-to-noise ratio. This subsection designs illumination strategies for dual-RIS-aided imaging systems to enhance the impact of the object on the received signal. We leverage the reconfigurable capability of RISs to provide diverse measurement patterns and jointly control the dual-RIS to bring synergistic benefits. 

\begin{figure*}
    \centerline{\includegraphics[width=2\columnwidth]{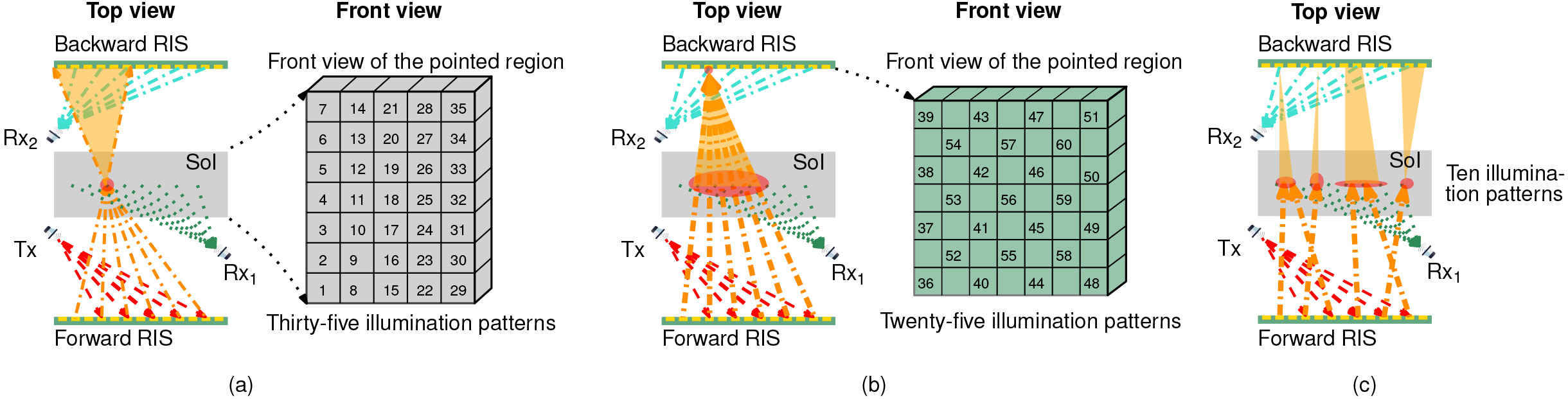}}
    \caption{llumination strategies. {{(a)}} The first type of illumination strategy comprises 35 illumination patterns, with the forward RIS beam focused on different cells of the SoI. The focused areas are marked by numbers from 1 to 35. {{(b)}} The second type of illumination strategy includes 25 illumination patterns, with the forward RIS beam focused on different cells of the backward RIS. The focused areas are marked by numbers from 36 to 60. {{(c)}} The third type of illumination strategy involves selecting 10 random illumination patterns.} 
    \label{tomography_illumination}  
\end{figure*}  

Specifically, three types of illumination strategies are designed to achieve imaging with a minimal number of measurements, as illustrated in Fig.~\ref{tomography_illumination}. These strategies integrate spotlight modes and multiplex measurement patterns from traditional illumination strategies. We devise an illumination strategy to enhance the reflected signal from the target, as well as another strategy aimed at maximizing the sensitivity of the target to the transmitted signal. Additionally, the third illumination strategy involves random radiations to enrich the enhanced multiplex measurement pattern. Three illumination strategis ensure a balance between spatial resolution and spatial observation scale. 

\subsubsection{Illumination Strategy \Rmnum{1} (Focusing on Reflection-mode Signals)} To maximize the perceptual capability of the reflected signals and improve spatial resolution, the 2 \text{m}$\times$2 \text{m}$\times$1 \text{m} SoI is divided into 35 cells, as illuminated in Fig.~\ref{tomography_illumination}{a}. The forward RIS beamforming patterns, essential for focusing on the 35 cells, are pre-designed using a greedy fast beamforming algorithm~\cite{9551980} and then sequentially employed to direct the electromagnetic waves towards each cell. The backward RIS's radiation method involves aggregating scattered signals corresponding to the 35 patterns of the forward RIS. By traversing the backward RIS units with the adaptive beamforming algorithm, the received signal power for $\text{Rx}_2$ is maximized, achieving the ``two-hop'' transmission. 

Using the above illumination strategy, when there are no reflectors in the SoI, $\text{Rx}_1$ receives the signal reflected from the environment, while $\text{Rx}_2$ receives a stronger signal due to beamforming. However, when there is a reflector in the cell where the beam is directed, $\text{Rx}_1$ will receive a stronger reflected signal, and $\text{Rx}_2$ will receive a weaker transmitted signal due to blocking. The primary objective of the first type is to improve the perceived signal-to-noise ratio and achieve the highest possible spatial resolution in each cell.

\subsubsection{Illumination Strategy \Rmnum{2} (Focusing on Transmission-mode Signals)} The second type of illumination strategy involves partitioning the 1.4 m $\times$ 1.4 m backward RIS into 25 cells, with the forward RIS sequentially employing beamforming to focus on each of these cells. This process enlarges the reflective perceptual space within the SoI, as depicted in Fig.~\ref{tomography_illumination}{b}, by multiplexing scene information into a small measurement area. Additionally, it caters to the perception needs of the backward RIS. In the reflective mode of the backward RIS, scattered signals are aggregated based on the 25 illumination patterns of the forward RIS, thereby maximizing the transmitted signal power for $\text{Rx}_2$ through beamforming. 

The beamforming approach facilitates extensive area scattering signal perception (integration) within the SoI, emphasizing the localized spatial information reuse capability compared to the first type of strategies. The area under measurement and multiplexing is highlighted in red in Fig.~\ref{tomography_illumination}{b}. By focusing on partial SoI information reuse, this strategy reduces the overall measurement count while also significantly mitigating the impact of abnormal noise~\cite{duarte2008single}.

\subsubsection{Illumination Strategy \Rmnum{3} (Focusing on Randomly Radiated Signals)} To fully exploit information multiplexing, we configure 10 randomly generated illumination schemes for the forward RIS, as illustrated in Fig.~\ref{tomography_illumination}{c}. These schemes involve radiation patterns created by randomly varying the on/off states of the RIS elements in different proportions. The backward RIS treats incoming waves as far-field plane waves and employs beamforming to converge signals at $\text{Rx}_2$. This strategy achieves overall information multiplexing with different weights for the SoI. 

In this framework, the system offers seventy groups of illumination patterns, leveraging the highly reconfigurable diversity of the RIS. By spatially altering waveforms, it efficiently reuses multiscale information from local to global levels. Combining the advantages of beamforming technology and multiplexing, the system provides an effective object perception strategy for RIS imaging. The perceptual advantages in reflection-mode and transmission-mode are fully explored, ensuring the spatial resolution of small objects, enhancing sensing accuracy and facilitating the fast acquisition of object information through multiplexing. 

\begin{figure*}[htbp] 
    \centerline{\includegraphics[width=1.9\columnwidth]{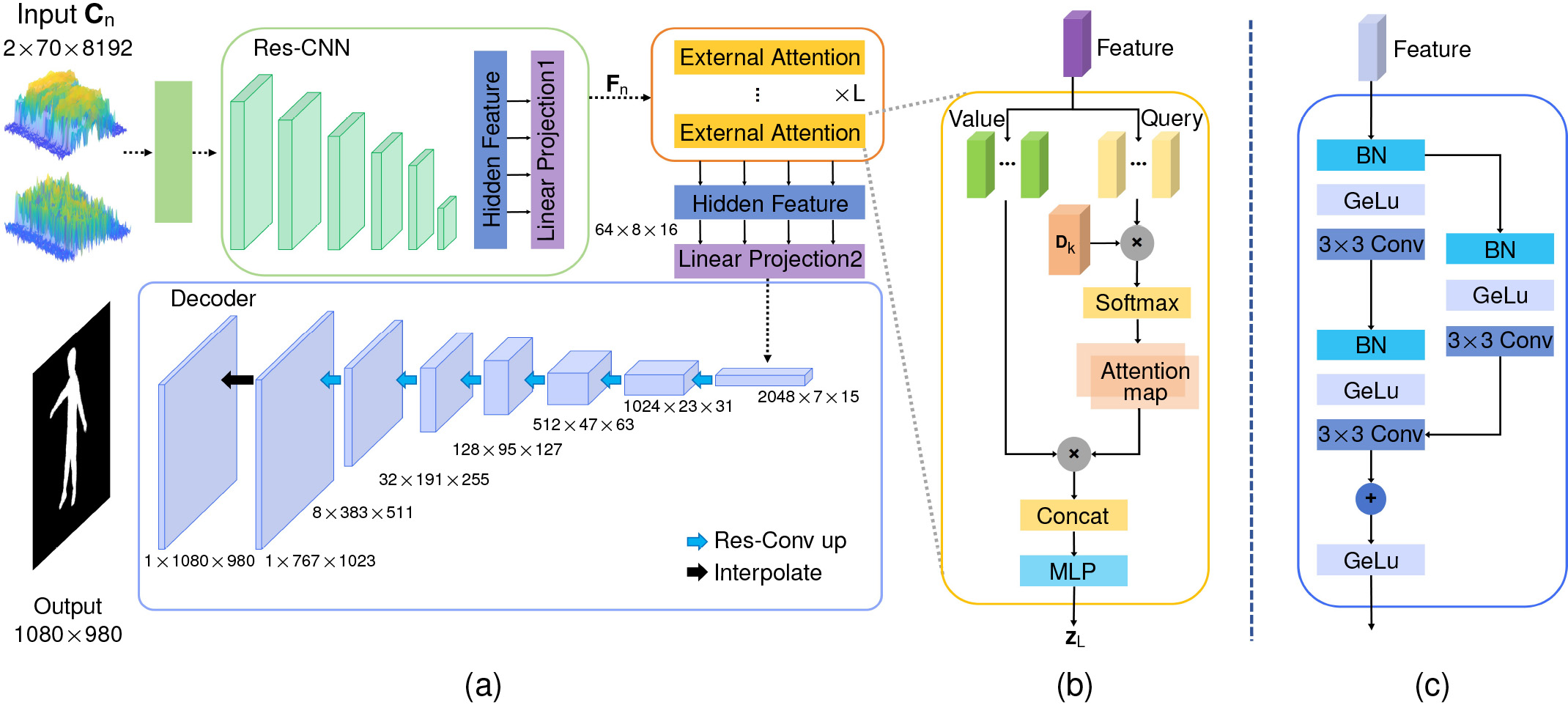}}
    \caption{Overview of the neural network PSD2ImageNet. {{(a)}} The architecture of PSD2ImageNet. It includes three main modules: an encoder, an external attention block, and a decoder. The encoder utilizes residual convolution blocks to extract features from power spectral density (PSD) data and convert them into a latent feature representation. The external attention block enhances the learning of deep feature representations. Finally, the decoder upsamples low-resolution human contour features to high-resolution contour features. {{(b)}} A schematic of the multi-head external attention layer. It is integrated into the encoder to capture long-range dependencies and refine feature representations. {{(c)}} A schematic of the Res-Conv block. It combines residual connections with convolution operations to facilitate feature extraction. } 
    \label{PSD2ImageNet_data_flow} 
\end{figure*} 

\subsection{Reconstruction Algorithm Design} 
We propose a straightforward yet efficient architecture, PSD2ImageNet, which learns a mapping rule from PSD data in the RF domain to a binary semantic image representing the human body in the spatial domain. The architecture is depicted in Fig.~\ref{PSD2ImageNet_data_flow}. Traditionally, CNNs are trained to encode RF data into high-level feature representations, which are then decoded back to the full spatial resolution. However, unlike existing approaches, our method introduces attention mechanisms into the encoder design through the use of external attention. 

With preprocessed PSD data from multiple Rxs, our approach is tailored to predict a binary semantic human body image. The data collection and preprocessing are presented in Section~\ref{S:Data Collection}. We denote the training dataset as ${\bf{D}}=\{ ({\bf{I}}_n, {\bf{C}}_n),n \in [1, N]\}$, where ${\bf{I}}_n \in \mathbb{R}^{1080 \times 980}$ and ${\bf{C}}_n \in \mathbb{R}^{2 \times 70 \times 8192}$ (2 Rxs, 70 illumination patterns and 8192 frequency points) are a pair of synchronised image frames and PSD data, respectively. Here, $n$ denotes the sampling human pose, and $N$ is the dataset size. 

Fig.~\ref{PSD2ImageNet_data_flow}{a} provides a deeper look into the network architecture, which consists of three key modules with a U-net architecture: \textbf{An encoder}, comprising residual convolution blocks, is responsible for extracting features from the PSD data and converting PSD data into a latent feature representation. We introduce \textbf{an external attention block} into the encoder to enhance the learning of deep feature representation. Notably, PSD2ImageNet is the first external-attention-based framework for RF imaging reconstruction. \textbf{An decoder}, responsible for upsampling low-resolution human contour features to high-resolution contour features with the Res-Conv blocks, plays a crucial role in the downstream task. 

\subsubsection{CNN-external-attention Hybrid as the Encoder} Initially, the input ${\bf{C}}_n \in \mathbb{R}^{2 \times 70 \times 8192}$ is reshaped into a $2 \times 560 \times 1024$ tensor, with every eight frequency points replaced along the illumination patterns axis, denoted as ${\bf{E}}_n$. The encoder then employs six residual convolution blocks for feature extraction, producing hidden layer features denoted as ${\bf{F}}_n$. Each Res-Conv block, illustrated in Fig.~\ref{PSD2ImageNet_data_flow}{c}, comprises three base convolution blocks, with rectified linear unit activations and convolutional layers following every batch normalization operation. 

\subsubsection{Multi-head External Attention Bottleneck} The encoder consists of $L$ layers of multi-head external attention (MEA) modules and multi-layer perceptron (MLP) blocks, as shown in Fig.~\ref{PSD2ImageNet_data_flow}{b}. These MEA modules capture long-range dependencies and refine feature representations, contributing to the network's overall performance~\cite{guo2022beyond}. \textbf{Multi-head external attention.} The EA computes attention between the input features ${\bf{z}}_{l-1}$ and an external memory unit ${\bf{M}}_k \in \mathbb{R}^{d}$, via:
\begin{equation}
    {\bf{A}} = \text{Softmax}(\text{MLP}({\bf{z}}_{l-1}) {\bf{M}}_k^\top),\quad 
    {\bf{D}}_k = \bf{A} \text{MLP}({\bf{z}}_{l-1}),
\end{equation}
where external memory unit ${\bf{M}}_k$ is a learnable parameter independent of the input, which acts as a memory of the entire training dataset. The attention map $\bf{A}$ is derived from this learned dataset-level prior knowledge, normalized similar to self-attention. The input features from $\bf{M}$ are then updated based on the similarities in $\bf{A}$.

\begin{figure*}
    \centerline{\includegraphics[width=2\columnwidth]{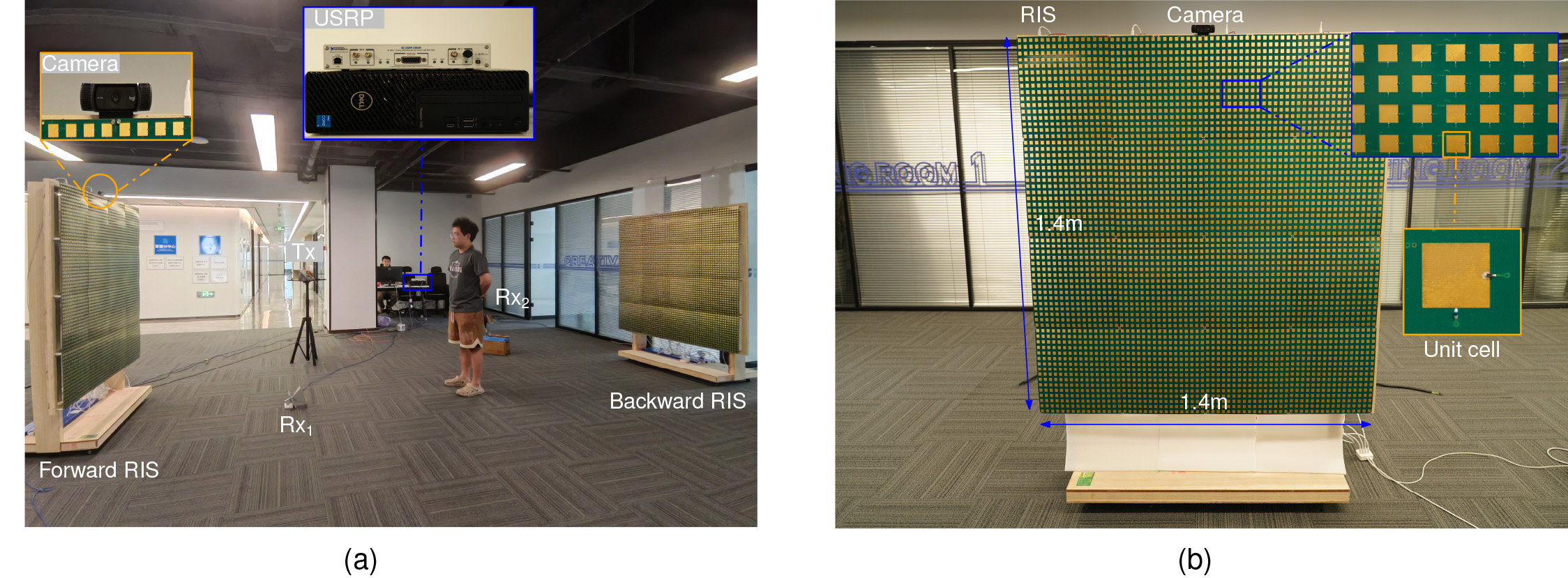}}
    \caption{{{(a)}} The dual-RIS-aided imaging system. It comprises signal transceiver equipment, two well-designed RISs, one Tx horn antenna, two Rx horn antennas, and a camera. {{(b)}} Employed RIS and its elements. The RIS employed in the system consists of $ K = 64 \times 64 $ units, operating at 5.8 GHz. {{(c)}} Collected Data. The collected data includes images and RF signals captured from various poses. (Top) The first row displays images of different poses. (Middle) The second row displays the reflected signals collected by $\text{Rx}_1$. (Bottom) The third row displays the transmitted signals collected by $\text{Rx}_2$.}
    \label{hardware_data}
\end{figure*}

The output of the $l$-th MEA layer is formulated as:
\begin{equation*} 
    {{\bf{h}}_i} = \text{EA}({\bf{z}}_{l-1},\bf{M}_k),\quad
    {\bf{z}}_l = \text{MLP}(\text{Concat}({{\bf{h}}_1},\cdots,{{\bf{h}}_H})),
\end{equation*}
where ${\bf{z}}_l $ is the encoded image representation, ${{\bf{h}}_i}$ is the $i$-th head, $H$ is the number of heads and ${\bf{M}}_k$ is the memory unit shared by different heads. In the bottleneck, the feature dimension and resolution remain unchanged. 

\subsubsection{The Decoder} The decoder is responsible for decoding the hidden features into the final contour output. Initially, the sequence of hidden features ${\bf{Z}}_n$ is reshaped to dimensions $64 \times 8 \times 16$. A $1 \times 1$ convolution operation is then applied to increase the number of channels from $64$ to $2048$. Multiple upsampling blocks are cascaded to progressively increase the resolution of the output. The output of the encoder is interpolated to the size $1080 \times 980$ and constrained to values between 0 and 1, producing the final contour output. 

PSD2ImageNet adopts a CNN-external-attention hybrid model for its encoder architecture, deviating from a pure Transformer approach. This hybrid model leverages CNN as a feature extractor to generate a feature map from the input data. Previous studies~\cite{petit2021u,cao2022swin} have demonstrated that this hybrid architecture often outperforms using a pure Transformer encoder~\cite{vaswani2017attention}. The Tab.~\ref{network:PSD2ImageNet} (see Appendix~\ref{A:network_architecture}) lists the detailed parameters of the feature extractor, providing additional context for the architecture and design choices.

\section{Experiment} \label{Experiment} 
\subsection{System Hardware}
The system setup of the proposed RIS-aided imaging system is illustrated in Fig.~\ref{hardware_data}{a}. It comprises signal transceiver equipment, two well-designed RISs, one Tx, and two Rxs. The dual-RIS-aided imaging system operates within the frequency of 5.8 GHz, utilizing a 160 MHz bandwidth. The universal software radio peripheral (USRP) 2954R, as illustrated in Fig.~\ref{hardware_data}{a}, is equipped with one Tx and two Rxs. The transmitter power of the signal is about 10 dBm. Both RISs share the same structure, measuring about 1.4 m$\times$1.4 m, with different functional roles. Each RIS consists of $ K = 64 \times 64 $ units, with the state of each unit controllable via on/off diodes, as illustrated in Fig.~\ref{hardware_data}{b}. The control signals for each unit's state are transmitted from the host computer to the microcontroller.

\subsection{System Setup and Signal Acquisition}
One of the RISs, designated as the forward RIS, collaborates with the Tx to modulate the signal scattering pattern, thereby imaging the SoI. We define the SoI as a space with dimensions of 2 m $\times$ 2 m $\times$ 1 m. The Tx, positioned 2 m to the left front of the forward RIS at a height of 1 m, emits signals directed towards the SoI. The SoI is situated in front of the forward RIS, while $\text{Rx}_1$ is positioned to capture signals reflected from the human body within the SoI. The other RIS referred to as the backward RIS, complements the forward RIS by capturing remaining signals that are diffracted, scattered, or reflected by the human body as the electromagnetic signal traverses the SoI. Subsequently, $\text{Rx}_2$ is situated 2 m to the right of the backward RIS to capture the transmitted signals reflected by the backward RIS. The entire RIS-aided imaging system is designed to be flexible and adaptable, suitable for installation on both sides of corridors or room walls to facilitate indoor human imaging surveillance, among other applications. 

\subsection{Data Collection} \label{S:Data Collection} 
We position the forward and backward RISs on opposite sides of the room and conduct imaging of the SoI located between the RISs. To assess the performance of the imaging system, we collect a dataset for evaluation on the aforementioned platform. We test 5 volunteers with heights ranging from $1.68$ m to $1.80$ m as subjects. 

We use the dual-RIS modulated with 160 MHz wideband signals, specifically LFM, for space point detection. We conduct 70 acquisitions of reconfigurable environments per posture, altering the position and pose after 70 acquisitions of an image-RF data pair to ensure diversity in the dataset. The system is equipped with two Rxs, and the total acquisition time for one posture is approximately one second. The dimension of the acquired wireless data PSD is $ 2 \times 70 \times 8192 $. Each wireless data set corresponds to an image data acquisition with dimensions of $ 1080 \times 960 $. In total, we collect 9613 data pairs.

\begin{figure}
    \centerline{\includegraphics[width=1\columnwidth]{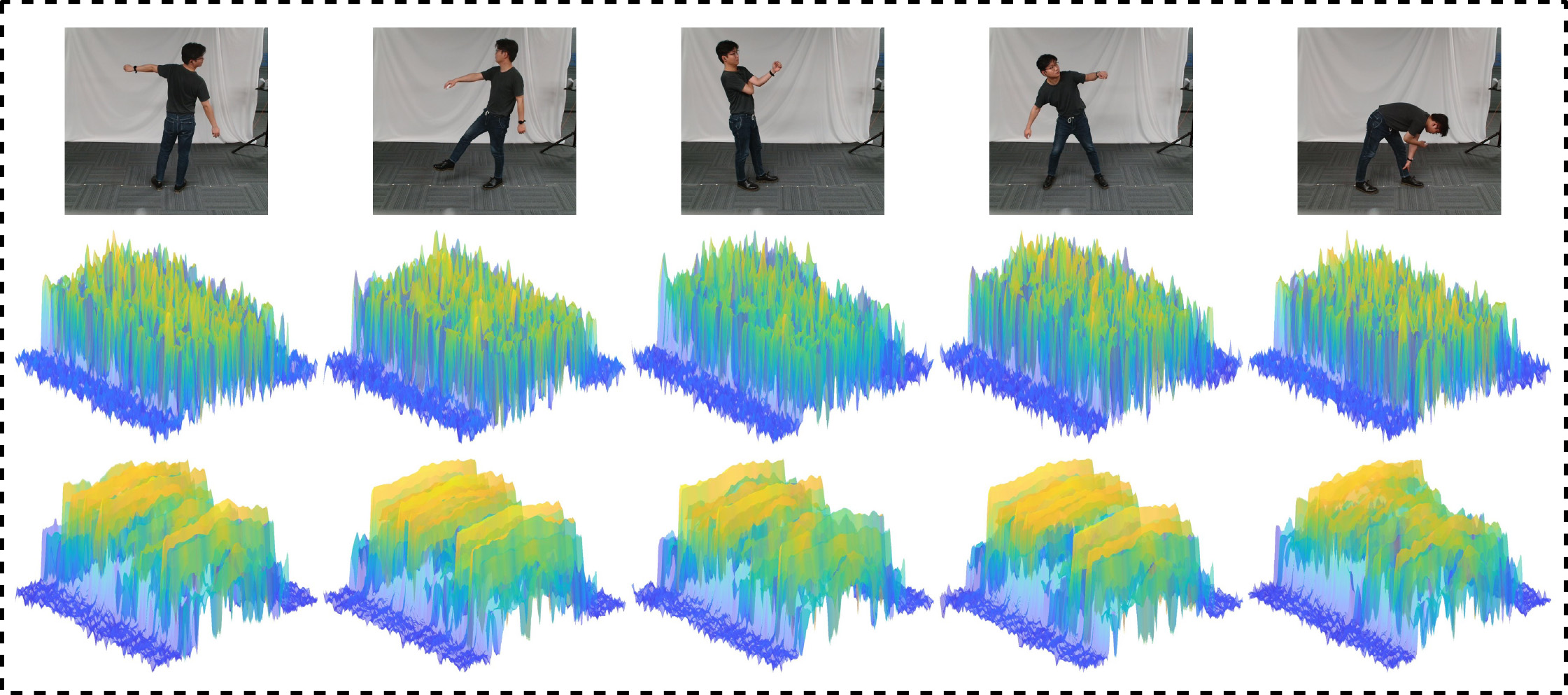}}
    \caption{Collected data. (Top) The first row displays images of different poses. (Middle) The second row displays the reflected signals collected by $\text{Rx}_1$. (Bottom) The third row displays the transmitted signals collected by $\text{Rx}_2$.}
    \label{F:data}
\end{figure}

We apply the median filter method for denoising to mitigate interference from white noise and device anomalies, as depicted in Fig.~\ref{F:data}. The image data is initially processed to acquire perceptual region portraits with a standardized dimensional size of $1080 \times 980$. Human contours are extracted using the publicly available segmentation network $\text{U}^2\text{net}$~\cite{qin2020u2}. 

\subsection{Training/Testing Protocols and Metrics} 
We randomly select $80 \%$ of the samples (7690 samples) to be our training set, and the remaining $20 \%$ (1923 samples) to be our testing set. The training and testing samples differ in the person's location and pose. We evaluate the performance of our proposed framework in its ability to detect body contours. $P(x, y)$, $G(x, y)$ denote the pixels of the prediction and the ground truth at position $(x, y)$, respectively. $G(x, y) \in \left\{ 0, 1 \right\}$ and $P(x, y)$ is the predicted probability of being a salient object. $H$ and $W$ are the height and width of the image, respectively. 

In evaluating the quality of the reconstruction image, we introduce five metrics to assess our model and the existing state-of-the-art algorithms: 

(1) \textbf{Structural similarity index measure (SSIM)}~\cite{wang2004image}. It is defined as: 
\begin{equation}
    SSIM=\frac{\left(2 \mu_{x} \mu_{y}+c_1\right)\left(2 \sigma_{xy}+c_2\right)}{\left(\mu_{x}^{2}+\mu_{y}^{2}+c_1\right)\left(\sigma_{x}^{2}+\sigma_{y}^{2}+c_2\right)},
\end{equation}
where $\mu_{x}$ and $\mu_{y}$ are the average pixel values of the reconstruction image and the ground truth, respectively. $\sigma_x$ and $\sigma_y$ are the variances, $\sigma_{xy}$ is the covariance between the reconstruction image and the ground truth. $c_1$ and $c_2$ are the constants used to stabilize the equation. Here, we set $c_1 = (k_1D)^2$, $c_2 = (k_2D)^2$, where $D$ is the dynamic range of intensity values. The SSIM value ranges from 0 to 1, where a value closer to 1 indicates that the reconstructed image has a more similar structure to the ground truth image.

(2) \textbf{MAE}. MAE evaluates the average per-pixel difference between a predicted saliency map and its ground truth contour.
\begin{equation}
    MAE = \frac{1}{W \times H} \sum_{x=1}^W \sum_{y=1}^H \left| P(x,y)-G(x,y) \right|.
\end{equation}

(3) \textbf{F-measure}. F-measure~\cite{margolin2014evaluate}, denoted as $F_{\beta}$, is an overall performance measurement and is computed by the weighted harmonic mean of the precision and recall:
\begin{equation}
    F_{\beta} = \frac{ \left( 1+{\beta}^2 \right) \times Precision \times Recall}{{\beta}^2 \times Precision + Recall},
\end{equation}
where $\beta$ is typically set to 1. 

(4) \textbf{Binary Cross Entropy (BCE)}. BCE~\cite{de2005tutorial} loss is widely used loss in binary classification and segmentation. It is defined as: 
\begin{equation}
    \begin{aligned}
    BCE = - \sum_{x=1}^H \sum_{y=1}^W &\left[ G(x,y) \log P(x,y) + \right.\\
                                                    &\left. (1-G(x,y)) \log(1-P(x,y)) \right].
    \end{aligned}
\end{equation}

(5) \textbf{IoU}. IoU~\cite{mattyus2017deeproadmapper} evaluates the overlap of the ground truth and prediction region. It is defined as: 
\begin{equation}
    IoU = \frac{\sum_{x=1}^W \sum_{y=1}^H P(x,y)G(x,y)}{\sum_{x=1}^W \sum_{y=1}^H [P(x,y)+G(x,y)-P(x,y)G(x,y)]}.
\end{equation}

The loss function, $L_\text{SSIM} + L_\text{MAE}$, supervises the structural consistency between the prediction and the ground truth during the training process, which is expressed as
\begin{equation}
    L_\text{loss} = L_\text{SSIM} + L_\text{MAE}= 1- \text{SSIM} + \text{MAE}. 
\end{equation}

\subsection{Implementation Details}
For the hybrid encoder design, we employ one external-attention layer with 16 heads and dropout set to 0.3. The learning scheme of the PSD2ImageNet is implemented using PyTorch, and the model is trained on one NVIDIA GeForce GTX 3090Ti with a batch size of 8. The network is trained for 100 epochs using the Adam optimizer with an initial learning rate of $1e^{-2}$, a momentum of 0.9 and a weight decay of $1e^{-4}$.

\section{Performance Analysis} \label{S:Performance analysis} 
\subsection{Imaging Results} 
The average SSIM between the output images generated by the well-trained PSD2ImageNet and the ground truth contours obtained from real optical images reaches a notably high value of 0.834 on the test set. Please refer to Appendix~\ref{A:loss} for more training details. The obtained output demonstrates the efficacy of our model in accurately reconstructing object contours. High-resolution imaging results are showcased in Fig.~\ref{result}, with additional results available in Fig.~\ref{reconstructed2} and Fig.~\ref{reconstructed3} (see Appendix~\ref{A:more_comparisons}). Experiments show that although imaging is feasible using either the reflection mode or the transmission mode with the assistance of RISs, combining the two modes can complement each other, yielding higher-quality images. These results qualitatively exhibit contours produced by the dual-RIS-aided imager. Notably, the contours of body locations, torsos, and legs are well-segmented with high-resolution, outperforming Wi-Fi-based imaging~\cite{Wang2019PersoninWiFiFP}. This level of performance is particularly beneficial for safety applications such as intrusion detection, detecting falls among the elderly, and identifying physical conflicts among individuals.

\begin{figure}
    \centering
    \subfigure{\includegraphics[width=1\columnwidth]{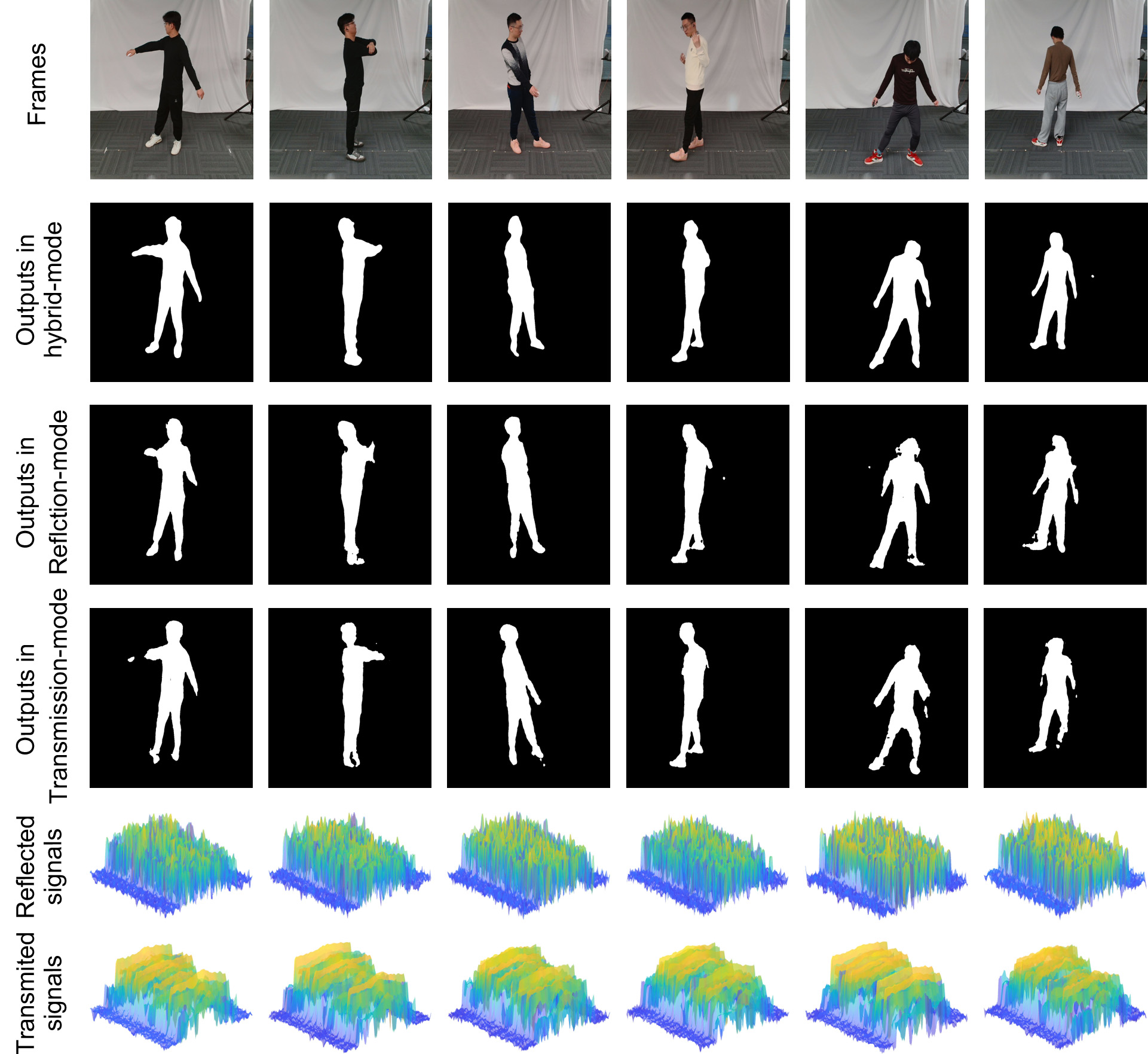}}
    \caption{Qualitative comparison of reconstructed human contours using different mode signals and the corresponding collected data. The highest quality of contour imaging is achieved by using hybrid-mode signals.}  
    \label{result}
\end{figure}

\subsection{Ablation Experiment}
This section elaborates on an ablation study to elucidate the impacts of the attention module, received signals in different modes, and various types of configurational diversities on reconstructing contour correspondence. 

\subsubsection{Impact of the Attention Module}
The experiment starts with training a baseline RF-Image model, serving as a reference, which excludes the attention module. The performance metrics for this baseline model are listed in the first row in Tab.~\ref{Attention Module}. Subsequently, the self-attention module is integrated, resulting in a noticeable enhancement in the metrics compared to the baseline, as indicated in the second row of Tab.~\ref{Attention Module}. Further study of the external attention mechanism is conducted~\cite{guo2022beyond}. The model's architecture includes two different memory units ($\text{M}_q$ and $\text{M}_v$). This paper proposes PSDImageNet with an external memory unit ($\text{M}_k$). PSDImageNet achieves the highest metric among the studied cases (fourth row in Tab.~\ref{Attention Module}). These outcomes suggest that the attention module effectively captures pertinent information regarding the body contours, and leveraging external attention notably improves the model's capability to reconstruct subtle details. 

\begin{table}
    \renewcommand{\arraystretch}{1.2} 
    \caption{Ablation Study of the Attention Module (\textsl{SA}: Self-attention~\cite{vaswani2017attention}, \textsl{EA-\uppercase\expandafter{\romannumeral1}}: External-attention (Q, V)))} 
    \label{Attention Module} 
    \centering 
    \begin{tabular}{lcccccc} 
    \hline 
        Method                          &   heads                 & SSIM                & IoU              &    MAE             & $\text{F}_\beta$ &   BCE             \\  \hline
        Res-CNN                         & $\setminus$             & 0.691               & 0.5689           &   0.054            &     0.72         &   4.612           \\  \hline
        \multirow{2}*{Res-CNN+SA} & 8  & \underline{0.831}   &\bf{0.733}        & \bf{0.032}  & \bf{0.844}       &   2.806           \\  
                                                             & 16 & 0.830               &\underline{0.7316}& \bf{0.032}  & \underline{0.843}&   2.800           \\  \hline
        \multirow{2}*{Res-CNN+EA-\uppercase\expandafter{\romannumeral1}}  & 8  & 0.704               & 0.571            &   0.056            &    0.724         &   4.470           \\ 
                                        & 16                      & 0.709               & 0.578            &   0.055            &    0.730         &   4.396           \\  \hline
        \multirow{2}*{\bf{PSD2ImageNet}}   & 8                       & \underline{0.831}   & 0.726            &\underline{0.033}          &    0.840         & \underline{2.637} \\ 
                                        & 16                      &\bf{0.834}           & 0.728            &\underline{0.033}          &    0.842         & \bf{2.535}        \\  \hline 
    \end{tabular}
\end{table}

\begin{figure}
    \centering
    \subfigure{\includegraphics[width=1\columnwidth]{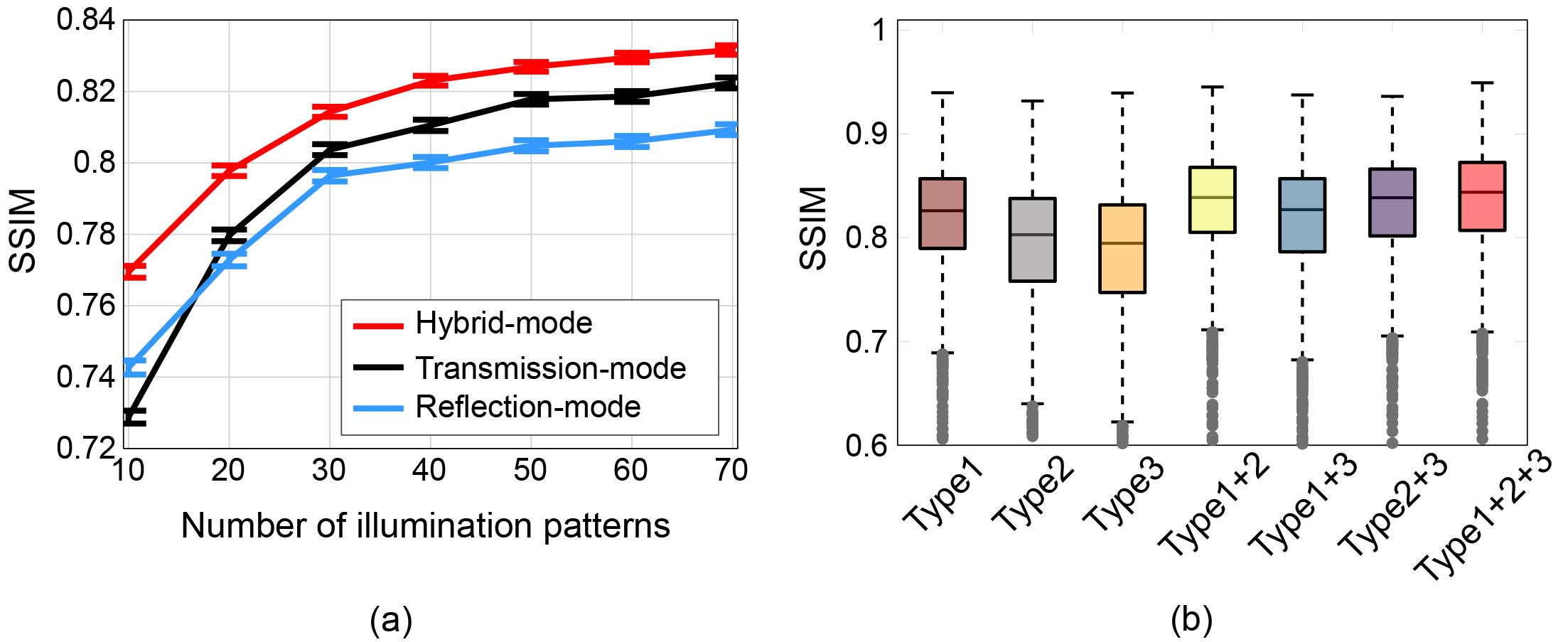}}
    \caption{Different modes of signals on image reconstruction performance analysis. {{(a)}} The average SSIM values are plotted as a function of the number of illumination patterns in hybrid-mode (red), reflection-mode (black), and transmission-mode (blue). It's evident that the highest resolution is obtained by reconstructing the human contour using hybrid-mode signals. As the number of illumination patterns increases, the imaging resolution improves. Furthermore, compared to reflection signals, signals in transmission-mode are more conducive to reconstructing clear contours. {{(b)}} Imaging metrics corresponding to different illumination strategies. The distribution of SSIM of the output images of 1923 data in the test set with respect to several types of illumination strategies is shown. It's observed that the highest imaging accuracy is achieved when three types of illumination methods work together. The next best performance is observed when the first two types of illumination methods work simultaneously. This suggests that a combination of different illumination strategies yields the best imaging results.} 
    \label{Reflection and transmission modes results and seven_types}  
\end{figure}


\subsubsection{Advantage of the Forward RIS and the Backward RIS}
The analysis of the ability of each RIS and hybrid-mode signals for object contour reconstruction reveals significant insights. When feeding only the single receiver signal of the collected data to the network for training and reconstruction, the SSIM results, illustrated in Fig.~\ref{Reflection and transmission modes results and seven_types}{a}, underscore the importance and necessity of both the forward and backward RIS for high-resolution imaging. Performance metrics for predicted imaging results are notably lower when only one type of Rx signals is fed. Therefore, it is concluded that the hybrid-mode can bring finer imaging accuracy. Additionally, it is noteworthy that the reconstructed images from the transmitted signal are cleaner compared to those from the reflected signal. Experimental results reveal that the transmission-mode has better perceptual advantages over the reflection-mode. 

\subsubsection{Effect of Different Illumination Strategies}
Classifying the illumination strategies into three categories, we analyze the distribution of SSIM of the outputs in the test set across various combinations of these strategies, as presented in Fig.~\ref{Reflection and transmission modes results and seven_types}{b}. The highest imaging accuracy is attained when three types of illumination methods work together. Following closely, the next best performance is observed when the first two types of illumination methods work simultaneously. This suggests that a combination of diverse illumination strategies leads to superior imaging accuracy, emphasizing the importance of leveraging multiple approaches for optimal results.

\begin{figure}
    \centerline{\includegraphics[width=1\columnwidth]{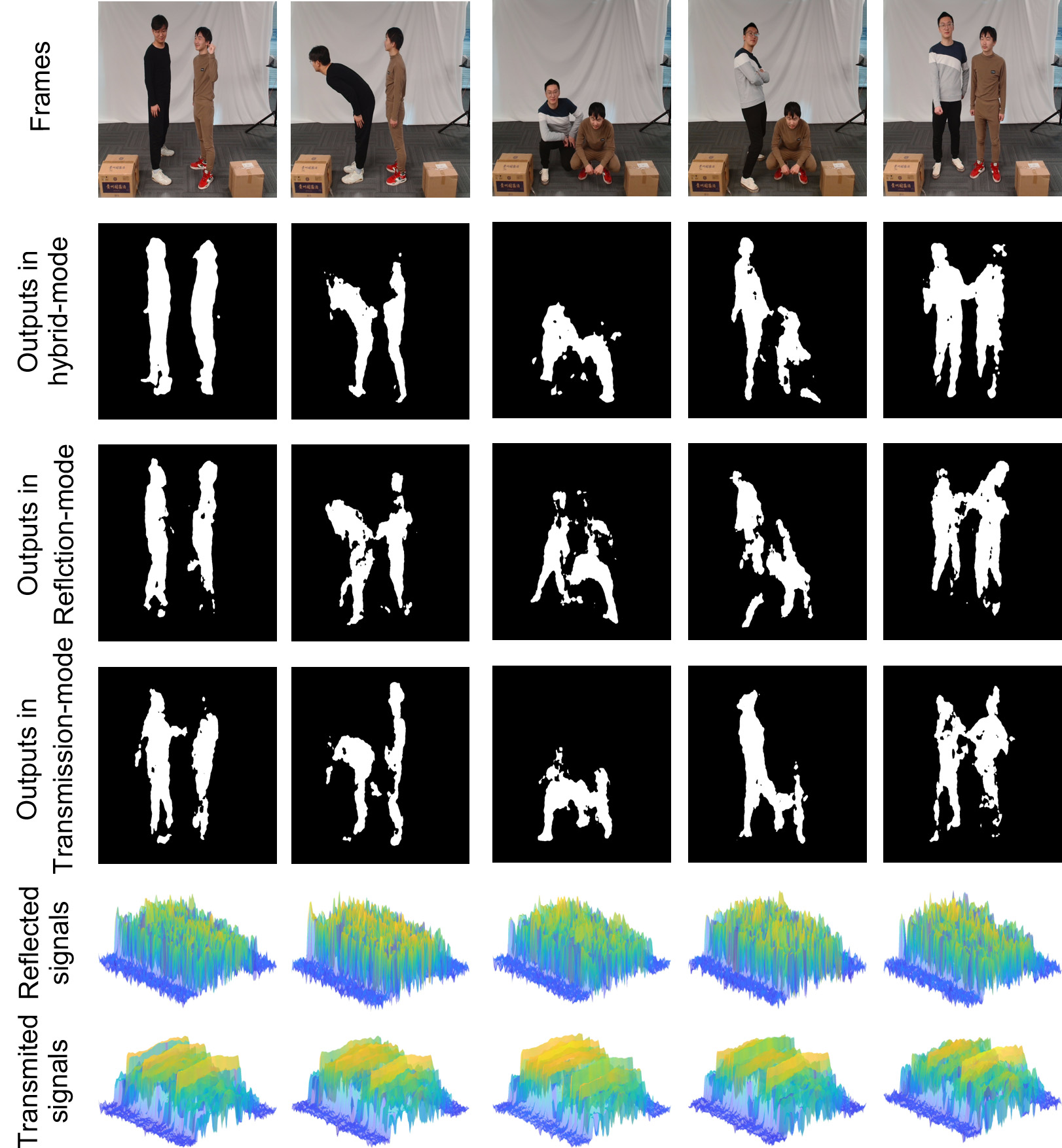}}
    \caption{The system's outputs in multi-target imaging. To demonstrate that our system can learn the physical relationships involved in electromagnetic inverse scattering, we use single-person data for training. The trained model is then employed to construct multi-target imaging. The reconstructed human contours are roughly shaped, meeting the requirements for determining the number of people and their postures.}
    \label{multiple persons}
\end{figure}

\subsection{Special Case Test}
In order to test the robustness of Dreamer, we conduct experiments in various special scenarios, including multi-target human imaging, line-of-sight obstruction, and low illumination conditions. Data collection and testing are performed under these conditions to assess the system's performance across different challenging scenarios. In addition, some limitations are discussed in Appendix~\ref{A:Discussions on limitations}.

\subsubsection{Human Imaging of Multiple Persons}
We evaluate the system's performance in multi-human imaging. Reconstructing multiple targets using wireless data, such as millimeter wave radar imaging~\cite{ding2023mi,9257482}, is a significant challenge. To demonstrate that our system can learn the physical relationship involved in electromagnetic inverse scattering, we use single-person data for training. The trained model is then employed to construct multi-target imaging. The imaging results depicted in Fig.~\ref{multiple persons} demonstrate the robustness and generalizability of our method. The reconstructed human contours are roughly shaped, with minor errors evident in some local limb areas. Further diversification of data and hard-example mining with multiple persons could enhance the results.

\subsubsection{Human Imaging with Obstructed View and Low Light}
\begin{figure}
    \centerline{\includegraphics[width=1\columnwidth]{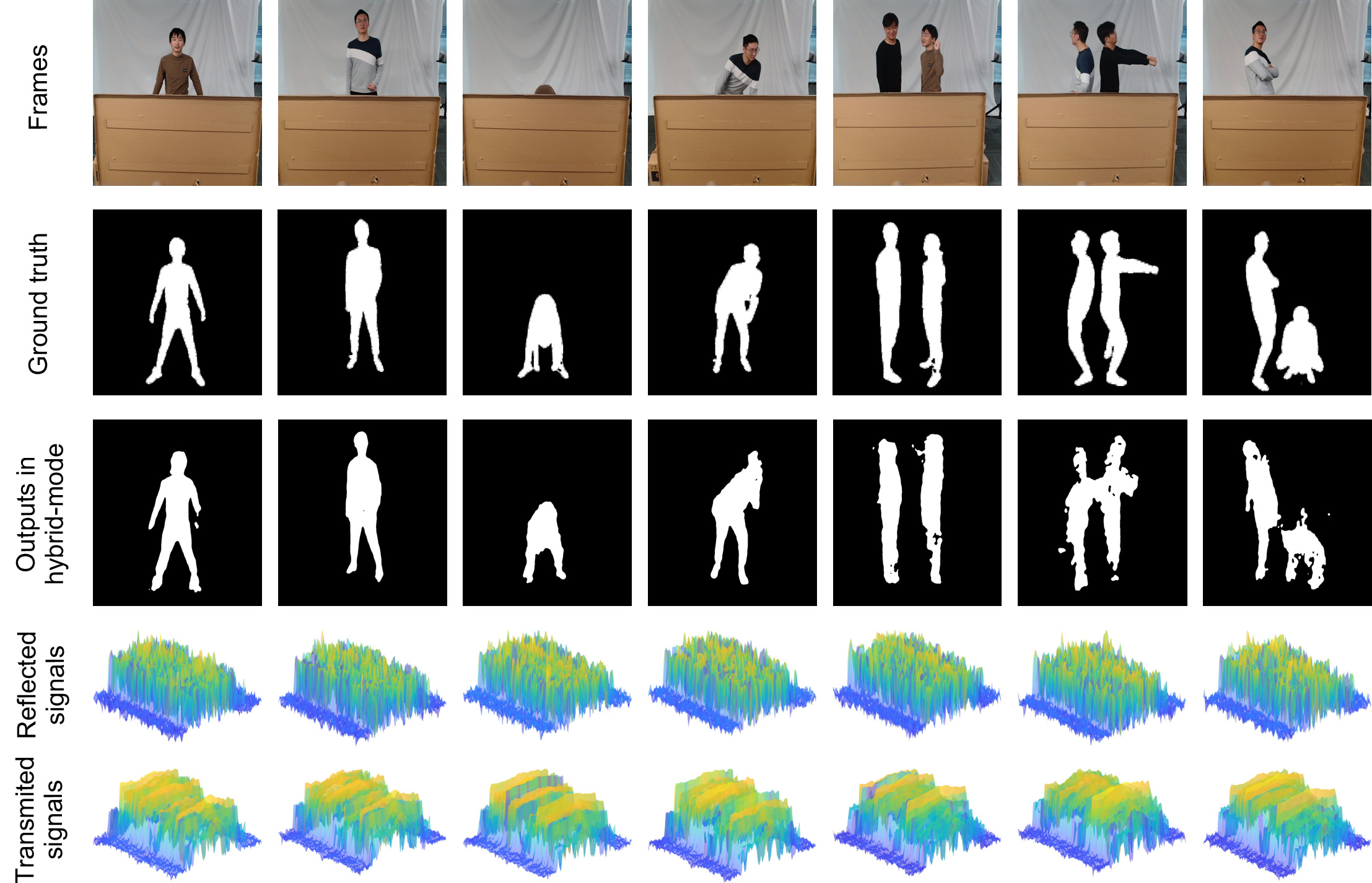}}
    \caption{The system's outputs with an obstructed view. The imaging results indicate that the proposed system is an effective complementary method to cameras when the view is obstructed.}
    \label{Obstructed}
\end{figure}

\begin{figure}
    \centerline{\includegraphics[width=1\columnwidth]{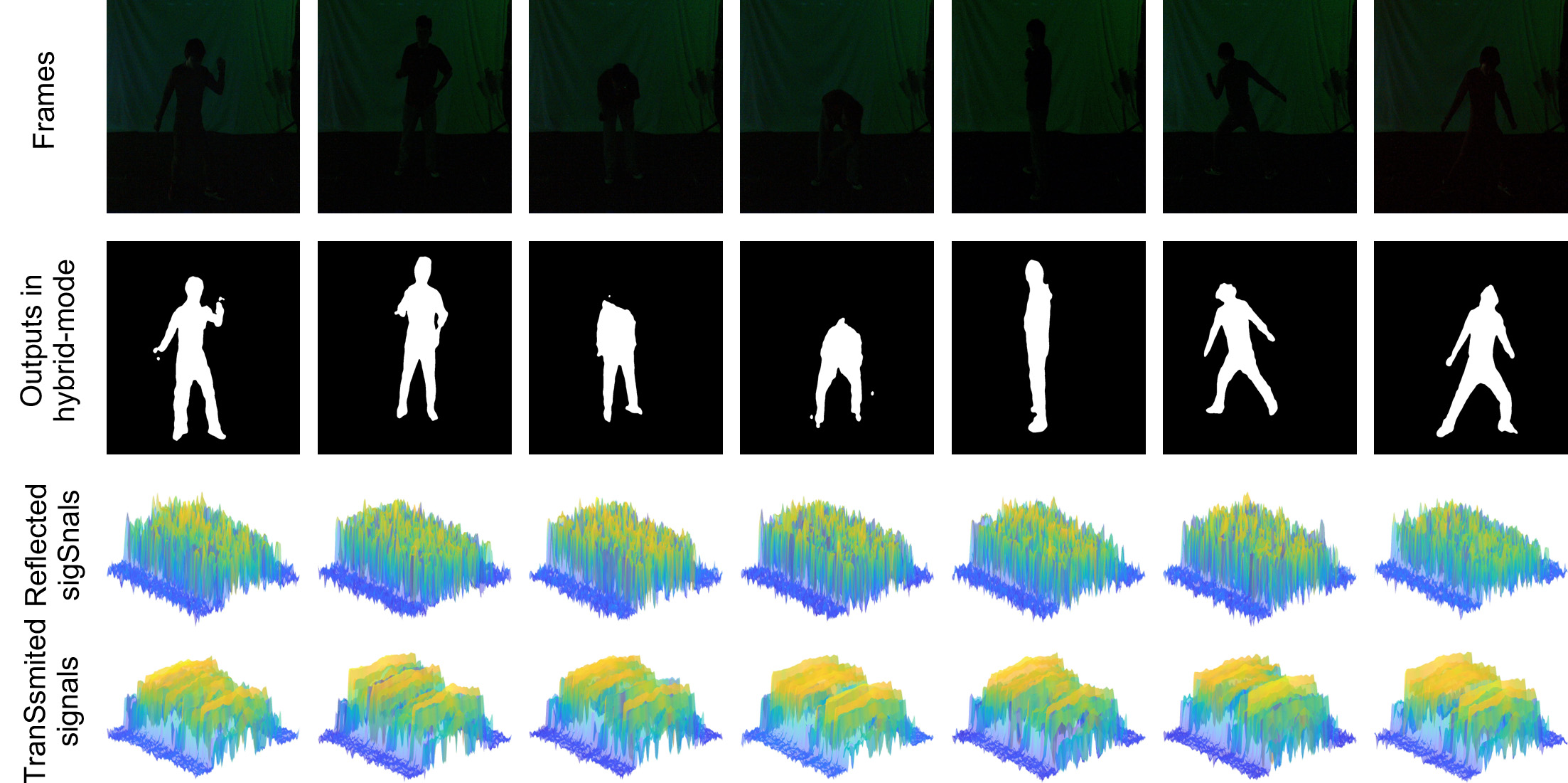}}
    \caption{The system's outputs in low light. The imaging results indicate that the proposed system is a good complementary method to cameras in low light.}
    \label{low light}
\end{figure}

The ability of RIS to perceive imaging through walls stems from its radio transmission capability. Fig.~\ref{Obstructed} illustrates that the system can image targets even when obstructed by a 5 cm thick cardboard. In low illumination conditions, RIS-aided imaging is not affected by the lighting environment compared to optical cameras. The results are shown in Fig.~\ref{low light}. Dreamer performs well under occlusion and low illumination conditions, making it a valuable auxiliary means for radio-assisted camera imaging when optical photography is limited.

\section{Conclusion} \label{S:Conclusion}
In this work, we have proposed the design and implementation of a dual-RIS-aided imager, the first system to achieve high-resolution RF imaging by exploiting the synergistic benefits of two RISs. Our proposed system, Dreamer, fully explores the perceptual advantages of hybrid transmission–reflection-mode. Experiments have shown that although imaging is feasible using either the reflection mode or the transmission mode with the assistance of RISs, combining the two modes can complement each other, yielding higher-quality images. We leverage the configurational diversities of dual-RIS and theoretically analyze how subjects affect the two complementary-mode signals through scattered object modeling. A prototype system is constructed and tested at 5.8 GHz in a typical indoor environment, yielding very promising results. We further develop a novel CNN-external-attention U-net neural network to translate PSD from RF signals into high-resolution images of human contours. It achieves remarkable performance with an SSIM exceeding 0.83, surpassing existing single-RIS-aided imaging systems. We anticipate that integrating the dual-RIS-aided imager into 6G and beyond wireless systems may stimulate the development of new perception algorithms and applications in the future. 

Compared with traditional methods, the proposed system fully leverages the reconfigurable and spatial degrees of freedom of RISs. By harnessing the synergy of distributed multiple RIS and transceiver antennas, new modes of imaging and sensing emerge. This approach can inspire reconfigurable environment perception and imaging tasks under large-scale RIS deployment, shaping the frameworks and theories of future multi-RIS collaborative imaging. Additionally, the scheme incorporates state-of-the-art deep learning network architecture, enabling modeling and reconstruction of complex electromagnetic propagation environments. Beyond 2D imaging tasks, this system can be expanded to reconstruct 3D targets and perform semantic perception tasks in 3D scenes. This not only redefines the concept of RISs but also lays the groundwork for the future metaverse and wireless perception.

\appendices

\section{The details of the network architecture} \label{A:network_architecture}
The details of the network architecture are given in Tab.~\ref{network:PSD2ImageNet}. The proposed network features three key modules with a U-net structure. The encoder, equipped with residual convolution blocks, extracts features from PSD data and transforms it into a latent representation. An external attention block is integrated to enhance deep feature learning. The decoder upscales low-resolution human contour features to high-resolution contours using Res-Conv blocks, crucial for the downstream task.

\begin{table}[htbp]
    \renewcommand{\arraystretch}{1}
    \caption{The network architecture of PSD2ImageNet (\textsl{Ch} represents the output channel, \textsl{s} represents the stride and \textsl{k} represents the kernel size)}
    \label{network:PSD2ImageNet}
    \centering
    \begin{tabular}{lcc} 
    \hline 
        \textbf{Input} & \multicolumn{2}{c}{$E_{n}\in R^{2\times560\times1024}$} \\ \hline \midrule  
        \multirow{2}*{Stage} &\multirow{2}*{Output Size}&Parameters\\
                           ~ &                          &$k_1 \times k_2$, ($s_1,s_2$), \textsl{Ch}\\ \hline
        \multirow{3}*{Res-Conv down1}     &\multirow{3}*{$8\times279\times511$ } &$3\times3,(2,2), 8$   \\  
                                  ~  &                                           &$3\times3,(2,2), 4$   \\ 
                                  ~  &                                           &$1\times1,(1,1), 8$   \\ \hline
        
       \multirow{3}*{Res-Conv down2}   &\multirow{3}*{$32\times139\times255$  }&$3\times3,(2,2), 32$      \\ 
                                 ~      &                                       &$3\times3,(2,2), 16$   \\ 
                                 ~      &                                       &$1\times1,(1,1), 32$   \\ \hline
        
        \multirow{3}*{Res-Conv down3}      &\multirow{3}*{$128\times69\times127$  }&$3\times3,(2,2), 128$      \\
                                 ~       &                                        &$3\times3,(2,2), 64$   \\ 
                                ~        &                                        &$3\times3,(1,1), 128$   \\ \hline

        \multirow{3}*{Res-Conv down4}      &\multirow{3}*{$512\times34\times63$  } &$3\times3,(2,2),512$      \\ 
                                ~           &                                     &$3\times3,(2,2), 256$   \\ 
                                 ~          &                                     &$3\times3,(1,1), 512$   \\ \hline

        \multirow{3}*{Res-Conv down5}      &\multirow{3}*{$1024\times16\times31$  }&$3\times3,(2,2), 1024$      \\ 
        ~                                   &                                     &$3\times3,(2,2), 1024$   \\ 
        ~                                   &                                     &$3\times3,(1,1), 1024$   \\ \hline

        \multirow{3}*{Res-Conv down6}      &\multirow{3}*{$2048\times7\times15$  } &$3\times3,(2,2),2048$      \\ 
                             ~             &                                       &$3\times3,(2,2), 2048$   \\ 
                            ~              &                                       &$3\times3,(1,1), 2048$   \\ \hline

        Linear Projection1              &$64\times7\times15$    &$1\times1,(1,1), 64$    \\ \hline
        \multirow{3}*{MEA}             &\multirow{3}*{$64\times7\times15$}    & number of heads = 16,\\
                                     ~  &                       & dropout =0.3, \\
                                     ~  &                       & hidden size =$64\times7\times15 $\\ \hline

        Linear Projection2  &$2048\times7\times15$  &$1\times1,(1,1), 2048$    \\ \hline
        
        \multirow{3}*{Res-Conv up1}        &\multirow{3}*{$1024\times23\times31$ }  &$3\times3,(2,2),1024$    \\ 
                                    ~       &                                       &$3\times3,(3,2), 1024$   \\ 
                                  ~         &                                       &$3\times3,(1,1), 1024$   \\ \hline

        \multirow{3}*{Res-Conv up2}         &\multirow{3}*{$512\times47\times63$ }&$3\times3,(2,2), 512$   \\ 
                                     ~      &                                     &$3\times3,(2,2), 512$   \\ 
                                  ~         &                                     &$3\times3,(1,1), 512$   \\ \hline

        \multirow{3}*{Res-Conv up3}        &\multirow{3}*{$128\times95\times127$}&$3\times3,(2,2), 128$   \\ 
                                    ~      &                                      &$3\times3,(2,2), 256$   \\ 
                                  ~        &                                      &$3\times3,(1,1), 128$   \\ \hline

        \multirow{3}*{Res-Conv up4}        &\multirow{3}*{$32\times191\times255$} &$3\times3,(2,2), 32$   \\ 
                                    ~      &                                      &$3\times3,(2,2),64$   \\ 
                                  ~        &                                      &$3\times3,(1,1), 32$   \\ \hline
                                  
        \multirow{3}*{Res-Conv up5}        &\multirow{3}*{$8\times383\times511$}  &$3\times3,(2,2), 8$   \\ 
                                    ~      &                                      &$3\times3,(2,2),16$   \\ 
                                ~          &                                      &$3\times3,(1,1), 8$   \\ \hline
                                
        \multirow{3}*{Res-Conv up6}        &\multirow{3}*{$1\times767\times1023$} &$3\times3,(2,2), 1$   \\ 
                                 ~         &                                      &$3\times3,(2,2), 4$   \\ 
                                 ~         &                                      &$3\times3,(1,1), 1$   \\ \hline
                                 
        Interpolate, clamp                 &$1\times1080\times980$                &  bilinear, min=0, max=1 \\ \hline
        \textbf{Output}                    & \multicolumn{2}{c}{$\hat{I}_n\in R^{1080 \times 980}$}          \\ \hline
    \end{tabular}
\end{table}

\begin{figure}[htbp]
    \centering
    \subfigure{\includegraphics[width=1\columnwidth]{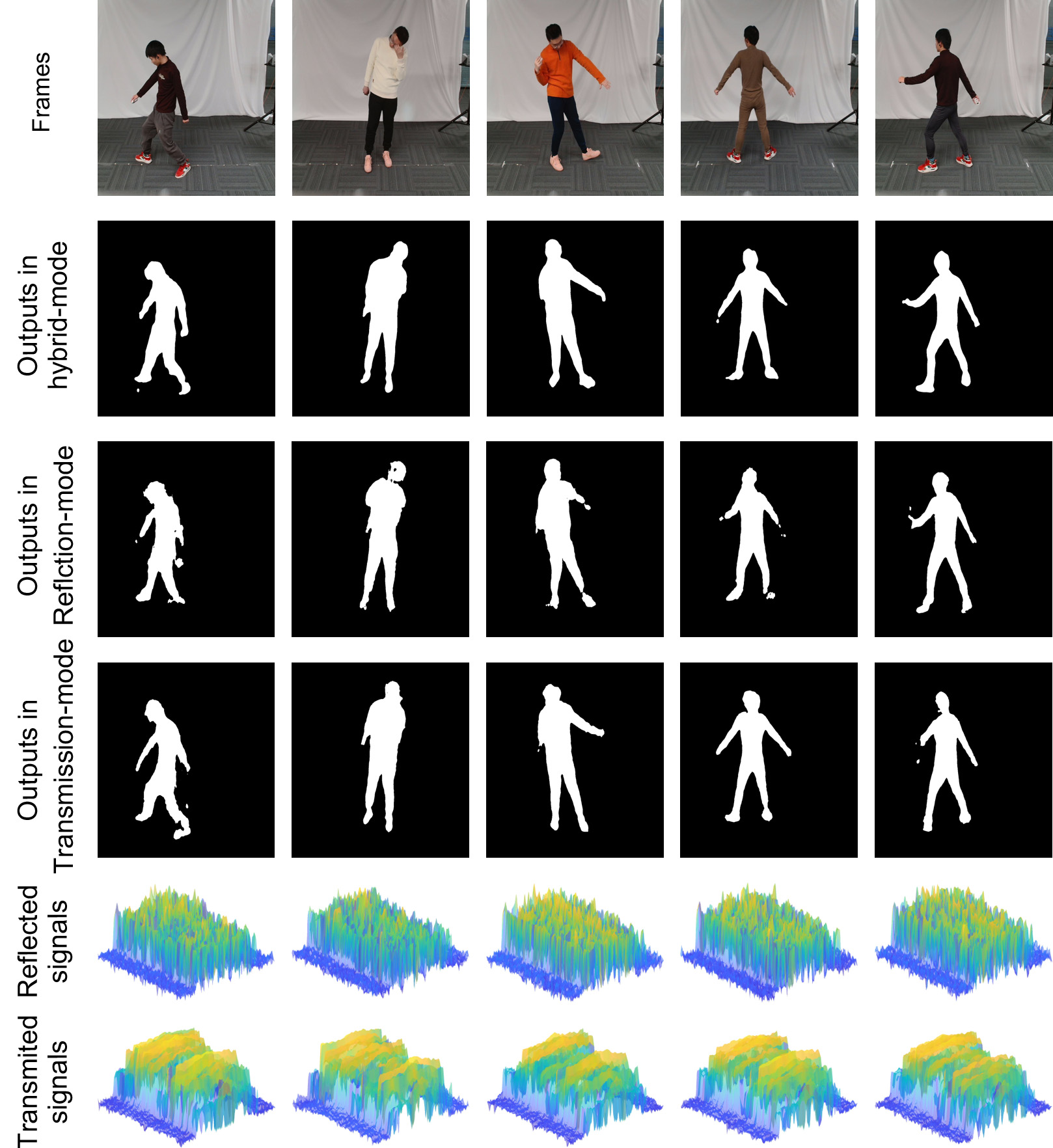}}
    \caption{More qualitative comparison of reconstructed human contours using different mode signals and the corresponding collected data. }
    \label{reconstructed2}
\end{figure}

\begin{figure}[htbp]
    \centering
    \subfigure{\includegraphics[width=1\columnwidth]{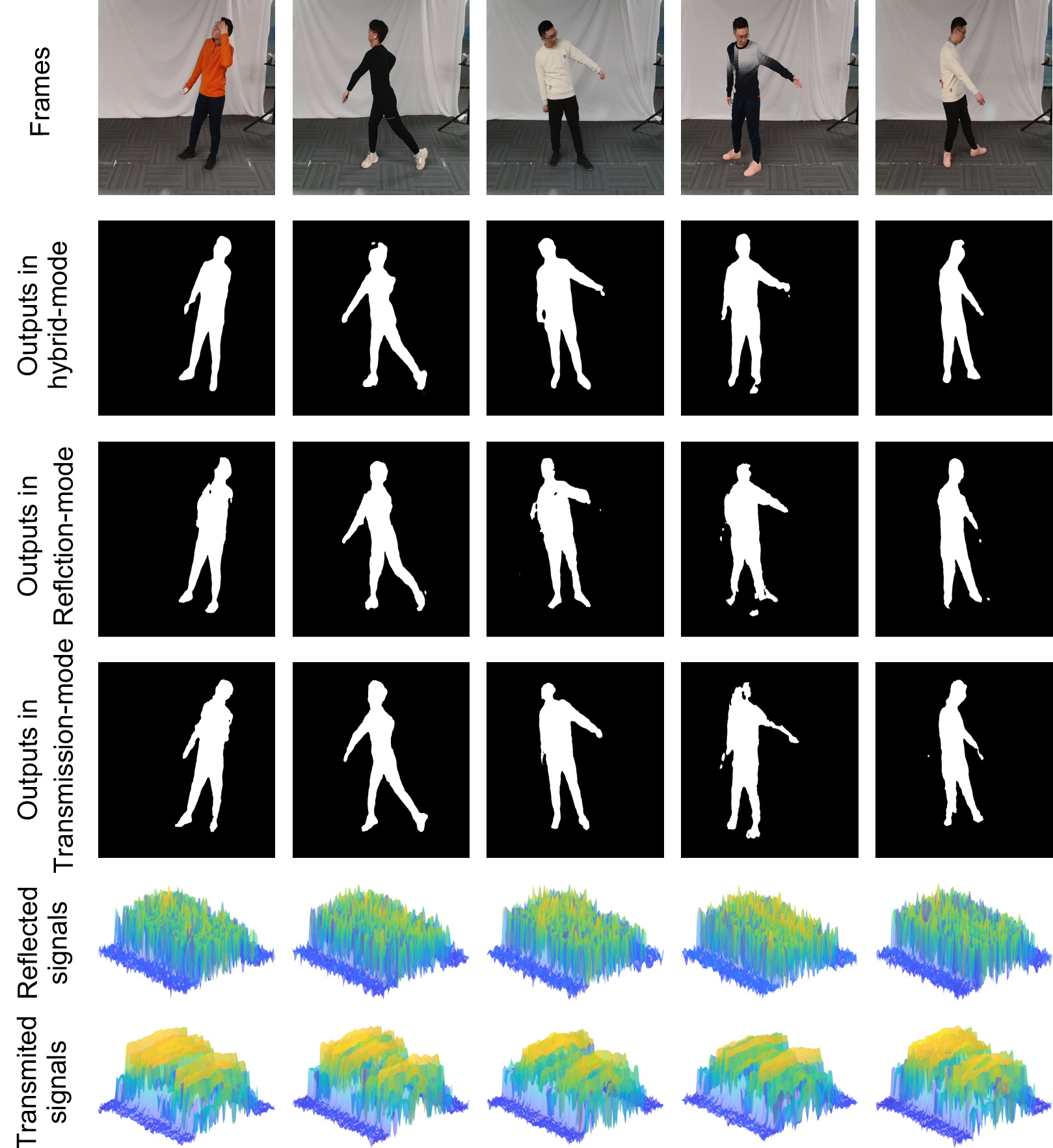}}
    \caption{More qualitative comparison of reconstructed human contours using different mode signals and the corresponding collected data. }
    \label{reconstructed3}
\end{figure}

\section{The more qualitative comparisons of reconstructed human contours}\label{A:more_comparisons}
As shown in Fig.~\ref{reconstructed2} and Fig.~\ref{reconstructed3}, more qualitative comparisons of reconstructed human contours are provided to show the capabilities of reconstructing object contours with different modes. The highest resolution is attained through reconstructing the human contour using hybrid-mode signals, showcasing qualitatively superior contours produced by the dual-RIS-aided imager.

\section{Discussions on Limitations} \label{A:Discussions on limitations}
\begin{figure}[htbp]
    \centering
    \subfigure[ ]{\includegraphics[width=1\columnwidth]{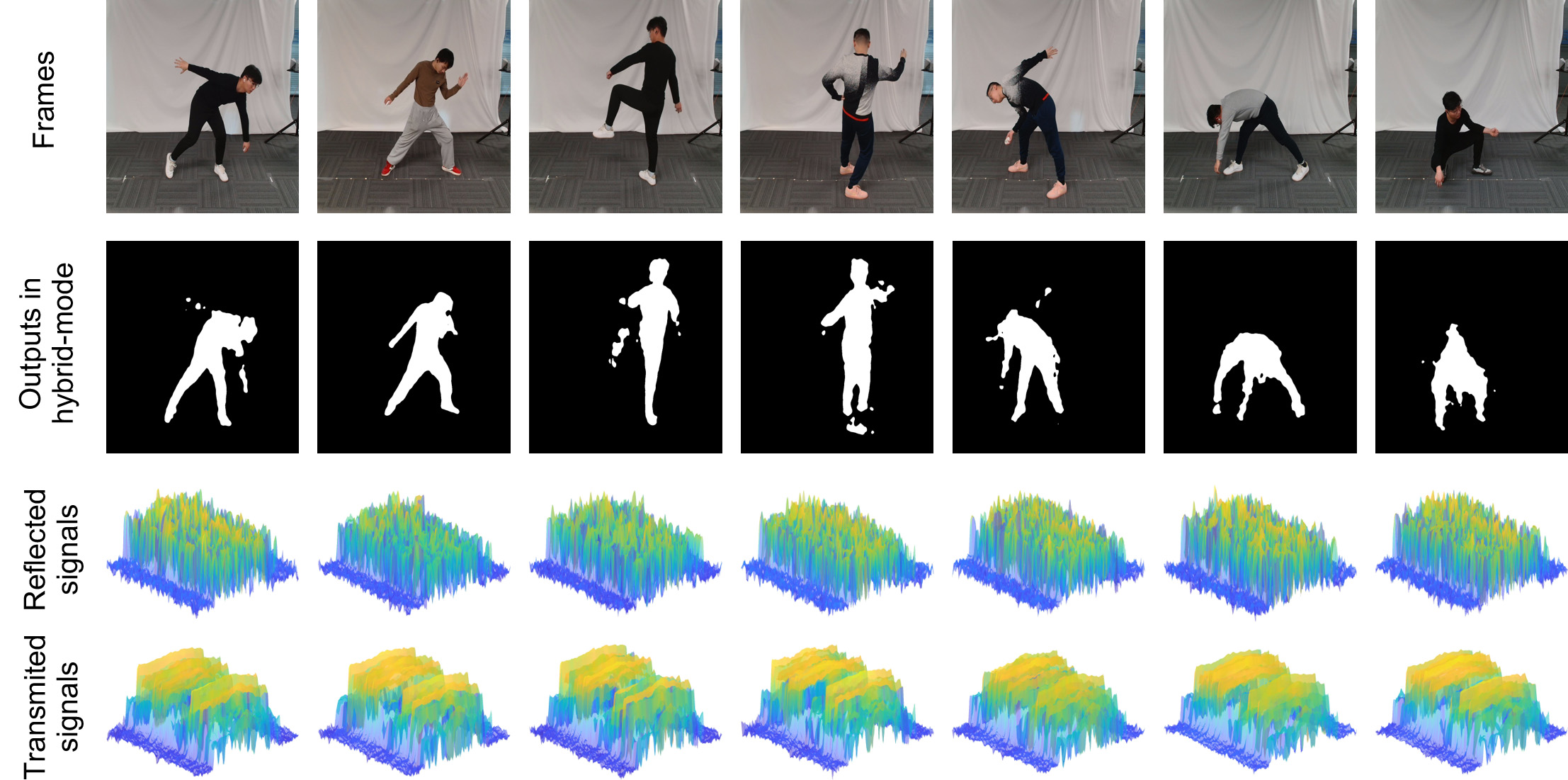}}
    \hspace{1in}
    \subfigure[ ]{\includegraphics[width=1\columnwidth]{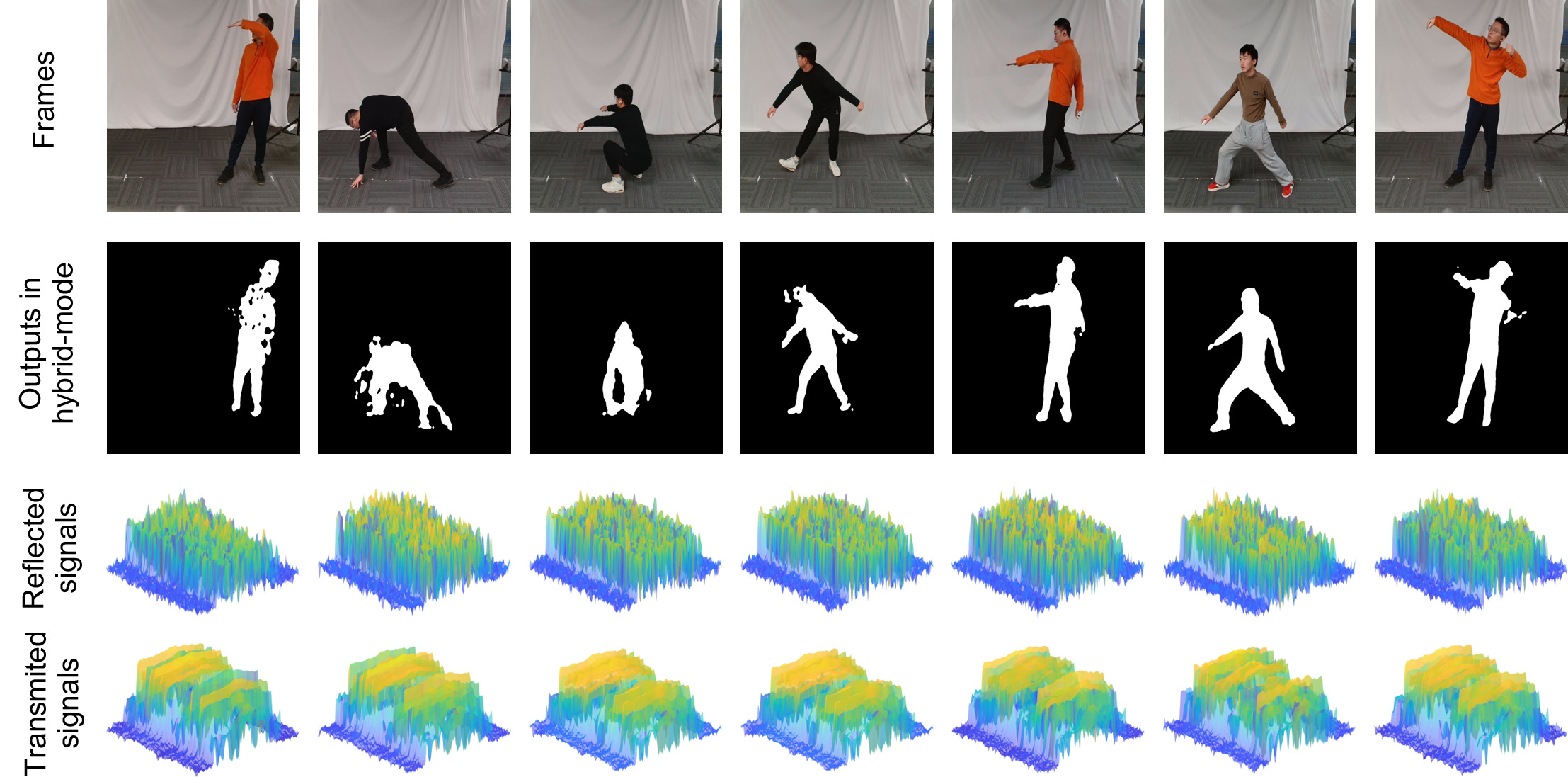}}
    \caption{Several failure cases in current results. (a) Failure cases where complex human body postures result in suboptimal imaging. (b) Failure cases where small limbs are absent.}
    \label{result3} 
\end{figure}

The failure cases illustrated in Fig.~\ref{result3} highlight certain limitations in our current results. Firstly, ambiguous complex poses (Fig.~\ref{result3}a) present challenges due to multiple-path radiation of electromagnetic signals caused by complex human body postures such as bending and kicking. The complexity of external perception information carried by electromagnetic waves makes it difficult for the system to accurately resolve high-resolution target contours. This issue could potentially be addressed by incorporating more spatial diversity to collect signals from additional directions. Secondly, the absence of small limbs (Fig.~\ref{result3}b) may occur due to small limbs being bypassed or mixed in electromagnetic waves because of diffraction or scattering effects. Improvements in results may be achieved through increased configurational diversity in data and hard-example mining. 

\section{The average loss of the training process}\label{A:loss}
\begin{figure}[htbp]
    \centerline{\includegraphics[width=0.8\columnwidth]{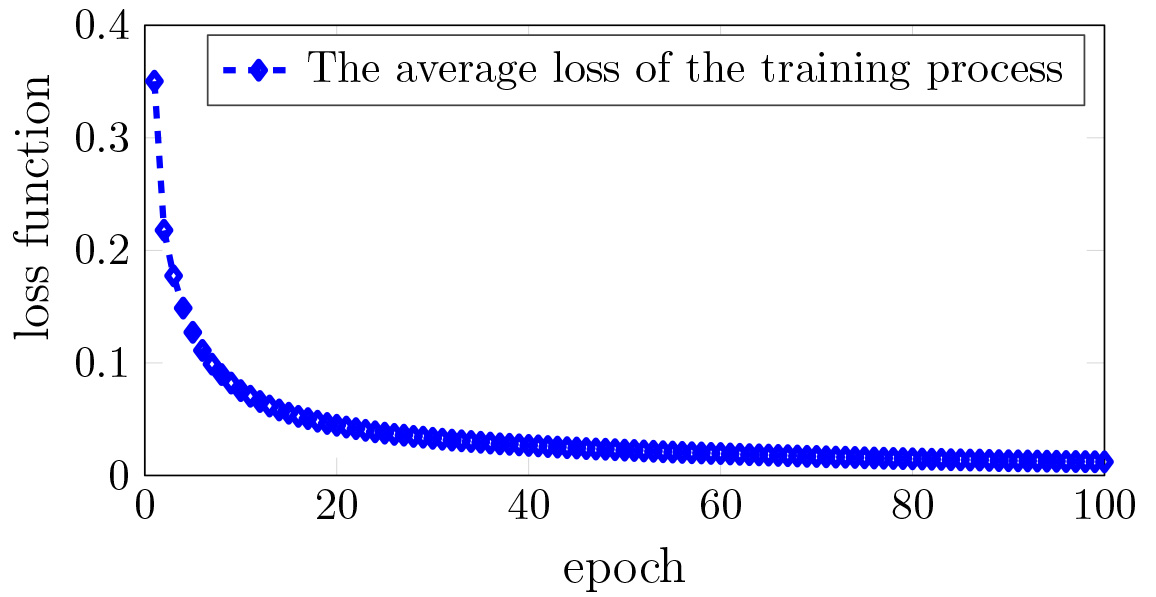}}
    \caption{The average loss of the training with 100 iterations.}
    \label{S5_loss}
\end{figure}

The training process's average loss is depicted in Fig.~\ref{S5_loss}. 

\bibliographystyle{IEEEtran}
\bibliography{Reference}

\end{document}